\documentclass[11pt]{article}

\usepackage{amsmath}
\usepackage{amsfonts}
\usepackage{amssymb}
\usepackage{hhline}
\usepackage{verbatim}

\usepackage{color}

\def\be{\begin{eqnarray}}
\def\ee{\end{eqnarray}}
\def\nn{\nonumber}

\def\FF{(-{\bf q^2t})}
\def\FFF{(- {\bf q^4 t})}
\newcommand{\blue}[1]{\textcolor{blue}{#1}}

\begin{document}


\hfill ITEP/TH-05/18

\hfill IITP/TH-03/18

\bigskip

\centerline{\Large{Are Khovanov--Rozansky polynomials consistent}}

\centerline{\Large{with evolution in the space of knots?}}

\bigskip

\centerline{{\bf A.Anokhina and A.Morozov }}

\bigskip

{\footnotesize
\centerline{{\it
ITEP, Moscow 117218, Russia}}

\centerline{{\it
Institute for Information Transmission Problems, Moscow 127994, Russia
}}
}

\bigskip

\centerline{\it ITEP, Moscow, Russia}

\bigskip

\centerline{ABSTRACT}

\bigskip

{\footnotesize
$R$-coloured knot polynomials for $m$-strand torus knots $Torus_{[m,n]}$
are described by the Rosso--Jones formula, which is an example
of evolution in $n$ with Lyapunov exponents, labelled by
Young diagrams from $R^{\otimes m}$.
This means that they satisfy a finite-difference equation (recursion)
of finite degree.
For the gauge group $SL(N)$ only diagrams with no more than $N$ lines
can contribute and the recursion degree is reduced.
We claim that these properties (evolution/recursion  and reduction) persist
for Khovanov--Rozansky (KR) polynomials, obtained by additional factorization
modulo $1+{\bf t}$, which is not yet adequately described in
quantum field theory.
Also preserved is some weakened version of differential expansion,
which is responsible at least for a simple relation between {\it reduced}
and {\it unreduced} Khovanov
polynomials.
However, in the KR case evolution is incompatible
with the mirror symmetry
under the change $n\longrightarrow -n$,
what can signal about an ambiguity in the KR factorization even for torus knots.
}

\bigskip

\bigskip

\section{Introduction}

Knot "polynomials" \cite{knotpols}
are the vacuum expectation values of Wilson loops
in $3d$ Chern--Simons theory \cite{CS}, perhaps, refined \cite{refCS}.
Since the theory is topological they depend only on the linking properties
of the loop ${\cal K}$ and do not change under smooth variations of
its embedding into $3d$ space ${\cal X}$.
For simply connected ${\cal X}=R^3$ or $S^3$ they are rational functions
(sometime, Laurent {\it polynomials}) of peculiar functions
$q = \exp\left(\frac{2\pi i}{g^{-2}+N}\right)$ and $A=q^N$ of the coupling
constant $g$ and the rank $N-1$ of the gauge algebra ${\cal G}=sl(N)$.
The shape of the polynomials depend on ${\cal K}$ and representation $R$
of ${\cal G}$.
The working definitions are in terms of the Reshetikhin--Turaev (RT) lattice theory
\cite{RT,RTmod} via Reidemeister-invariant convolutions
of quantum ${\cal R}$-matrices at the nodes of the $2d$ knot diagrams ${\cal D}$
and of Khovanov--Rozansky (KR) cohomological calculus \cite{Kh}--\cite{Ano}
on the hypercubes, associated with ${\cal D}$.
The central question in Chern--Simons theory is description of these
rational functions on the (discrete) space of knots or knot diagrams --
and it is a direct generalization of the similar question about $2d$ conformal blocks
\cite{CFT}, defined on the space of the 3-valent Feynman diagrams.

The first step in the story is of course the choice of parametrization of
the space itself.
Knot diagrams are often considered as glued from braids, and each $m$-strand braid has
a natural decomposition into an ordered (noncommutative) product of
the braid-group generators:
\be
{\cal B}\{n\} = \tau_1^{n_{1|1}}\ldots \tau_{m-1}^{n_{1|m-1}}\,
\tau_1^{n_{2|1}}\ldots \tau_{m-1}^{n_{2|m-1}}\,\ldots
\label{braid}
\ee
The first task in the theory of knot polynomials is to describe their dependence
on any of the integer-valued parameters $n$.
This approach is called {\it evolution method} \cite{DMMSS}-\cite{Arthdiff},
because as a function of $n$ the polynomial,
{obtained by the modified RT method} \cite{RTmod}, is a sum of Lyapunov-like exponents:
\be
{\cal P}_R^{\,{\cal K}(n)} = \sum_{Q\in R^{\otimes m}} C_{R,Q}\cdot \lambda_Q^n
\label{evo}
\ee
over the Young diagrams $Q$ in the product $R^{\otimes m}$
with $n$-independent coefficients $C$.
In other words, the polynomial satisfies a finite difference equation (recursion) in $n$,
\be
\left(\prod_{Q\in R^{\otimes m}} \hat\nabla_{\lambda_Q}^{(1)} \right)
{\cal P}_R^{\,{\cal K}(n)} = 0
\label{basrec}
\ee
which can be integrated to a system
\be
\Big(\prod_{\stackrel{Q\neq Q_0}{Q\in R^{\otimes m}}}
\hat\nabla_{\lambda_Q/\lambda_{Q_0}}^{(1)} \Big)
\left(\lambda_{Q_0}^{-n}\cdot {\cal P}_R^{\,{\cal K}(n)}\right) =
C_{R,Q_0}
\cdot \prod_{\stackrel{Q\neq Q_0}{Q\in R^{\otimes m}}}
\left(1-\frac{\lambda_Q}{\lambda_{Q_0}}\right)
\label{basrec1}
\ee
Here we use the finite difference operator
\be
\hat\nabla^{(m)}_x F_n = F_n-x\cdot F_{n-m}
\ee
and do not divide by $1-x$ to simplify most formulas below.
Also, we define the shift factor in $\nabla_x^{(m)}$ as $x$ rather than $x^m$,
because this simplifies formulas for superpolynomials.

If some $p$ of the expansion coefficients $C$ are vanishing,
the degree of the equation is actually reduced by $p$.
In particular, this happens for $N<m|R|$, because all the
diagrams $Q$ with more than $N$ lines do not contribute.
There are also some ``accidental'' omissions for HOMFLY polynomials
(like $C_{[1],[22]}$ in the case of torus $H^{[4,n]}$),
but they are usually lifted after ${\bf t}$-deformation to
superpolynomials.

While evolution recursion is a direct consequence of RT approach,
it is not yet so evident for KR method
(see the first paper 
in \cite{DM} for a tedious derivation in the simplest example).
Thus it is a meaningful question,
{\bf if KR polynomials satisfy a recursion (\ref{basrec1}),
and if its degree is further reduced at small $N$?}

In the present paper we analyse the available list of KR polynomials
for the simplest {\it torus knots} with $m=2,3,4,5 
$
in the fundamental
representation $R=[1]=\Box$ and observe
that the answer is "yes":
not only recursion is true, but its degree drops down at $N<m$,
exactly as it does for the corresponding superpolynomials --
where this followed from the super-Rosso--Jones formulas of \cite{DMMSS}.

If one assumes
the recursion, then it provides an answer for infinitely
many KR polynomials with all positive integer $n$
once the first few are known -- what is a major achievement by itself,
given the complexity of direct calculations in KR formalism.

However, as  already observed in \cite{Sat},
for some knot families the evolution formulas for superpolynomials
provide a unified description
only for positive values of the evolution parameter $n$,
while fail to provide pure positive or pure negative polynomials for negative $n$.
In this paper we find that the same happens to KR polynomials even for the
torus family with more than two strands.
In result, these functions of $n$ do not respect the mirror
symmetries $n\longrightarrow -n$, at least in the usual way:
while
\be
{\cal P} 
^{[m,-n]}({\bf a},{\bf q},{\bf t})
={\cal P} 
^{[m,n]}({\bf a}^{-1},{\bf q}^{-1},{\bf t}^{-1})
\label{invn}
\ee
for torus superpolynomials from \cite{DMMSS},
\be
{\cal K}^{[m,-n]}(N, {\bf q},{\bf t})
\neq {\cal K}
^{[m,n]}(N, {\bf q}^{-1},{\bf t}^{-1})
\label{Knoninvn}
\ee
for their KR counterparts, i.e. the left hand sides are understood as analytic
continuations from positive $n$ to negative, provided by the evolution formulas.
Despite mirror symmetry is not a true symmetry of knot theory,
associated link diagrams are not Reidemeister-equivalent, this is still an
unpleasant feature, signalling about some kind of non-analyticity of the
KR reductions in the space of knots (or, perhaps, about non-adequate
coordinatisation of this space by parameters like $n$).
Of course, the problem (\ref{Knoninvn}) disappears in cases when KR reduction
is trivial --  exhausted by just a substitution ${\bf a}={\bf q}^N$ in the
superpolynomial, like it happens for reduced 2-strand $(m=2)$ torus
polynomials at $N=2$.
For the same reason mirror symmetry survives
in the 3-strand  $(m=3)$ case for reduced torus
polynomials at $N=2,3$ and for the unreduced ones at $N=2$.

As to another symmetry, acting on the family of knot polynomials,
$[m,n]\leftrightarrow[n,m]$, which is topological and relates Reidemeister-equivalent
link diagrams, it is respected by our evolution formulas:
both are true,
\be
{\cal P}^{[m,n]}({\bf a},{\bf q},{\bf t})={\cal P}^{[n,m]}({\bf a},{\bf q},{\bf t})
\label{exchmn}
\ee
and
\be
{\cal K}^{[m,n]}(N,{\bf q},{\bf t})={\cal K}^{[n,m]}(N,{\bf q},{\bf t})
\label{Kexchmn}
\ee

\section{Recursion from Rosso--Jones formula for torus knots}

The simplest example of (\ref{evo}) appeared in the case of torus knots,
when the knot diagram consists of a single $m$-strand braid (\ref{braid})
of a very special form: with all non-vanishing $n_{a|i}=1$,
\be
{\cal B}^{[m,n]} = (\tau_1 \ldots \tau_{m-1})^n
\ee
One gets a knot when $n$ and $m$ are coprime, otherwise there is a link
with $gcd(m,n)$ components.
For $m$-strand torus knot HOMFLY polynomial is given by the Rosso-Jones evolution formula
\cite{RJ,DMMSS}
\be
^F\!H_R^{[m,n]}(A,q) = F_R^n \sum_{Q\in R^{\otimes m}} C_Q \cdot
q^{-\frac{2n}{m} \varkappa_Q}
\cdot D_Q(A,q)
\label{torusHOMFLY}
\ee
where $D_Q={\rm Schur}_Q^* :={\rm Schur}_Q\left\{p_k=\frac{A^k-A^{-k}}{q^k-q^{-k}}\right\}$
are quantum dimensions of representations $Q$,
given by the standard hook formula,
the eigenvalues of cut-and-join (actually, Casimir) operator are
$\varkappa_Q = \sum_{(i,j)\in Q}(i-j)$ and the integer-valued coefficients $C_Q$
are defined by the Adams rule
\be
{\rm Schur}_R\{p_{mk}\} =  \sum_{Q\in R^{\otimes m}} C_Q \cdot
{\rm Schur}_Q\{p_k\}
\label{AdamsHOMFLY}
\ee
Since the only dependence on $n$ is in the power of $q$,
this HOMFLY satisfies a homogeneous finite-difference equation
\be
\prod_{ {Q\in R^{\otimes m}}}
\hat\nabla^{(m)}_{\! q^{-2 \varkappa_{_Q}}\cdot F_{_R}^m}
H_R^{[m,n]} = 0
\label{heqHOMFLY}
\ee
where
and the degree of the difference equation is naively the number of Young diagrams
$Q\in R^{2m}$.
This fact is independent on the actual value of the coefficients $C_Q$,
i.e. from the point of view of the equation, they are the free parameters,
parametrising its solution.
In this sense the equation does not provide too much information about the
HOMFLY polynomials, still it reflects an important property -- evolution rule --
of the torus {\it family} with particular number $m$ of strands.

However, when some additional information is available, the degree of the
equation can be decreased.
For example, if the spectrum of Casimir eigenvalues $\varkappa_Q$ is degenerate,
i.e. if two or more of them coincide, the corresponding $Q$ should appear only once
in the product in (\ref{heqHOMFLY}).
{\it Such} degeneracy is often preserved by the ${\bf t}$-deformation.

If instead some of the coefficients $C$ are vanishing,
the corresponding $\nabla$ can be omitted
and the degree of the difference equation gets lower.
The lift from HOMFLY to superpolynomials, however,
usually lifts such degenerations: even if $C$ was vanishing for HOMFLY,
its ${\bf t}$-deformation does not, see examples below.

Another story is when vanishing are some dimensions $D_Q$.
For $A=q^N$ contributing are only $Q$ with no more than $N$ columns,
because otherwise $D_Q(q,q^N)=0$.
The order of the difference equation is reduced accordingly.
In particular, for $N=2$ $R$ and $Q$ are symmetric representations
(two-column representations of $sl_2$ are equivalent to the single-column ones).
{\bf The main claim  of the present paper will be that {\it this} reduction
is preserved by the ${\bf t}$-deformation -- even after additional
reduction from super- to KR polynomial.}
Still, in this section we  continue with HOMFLY.

If one adjusts the framing "phase" $F_R$ so that it
cancels the contribution from
one of the $\varkappa_Q$, $F_R=q^{\frac{2}{m}\varkappa_{Q_{_0}}}$,
then (\ref{heqHOMFLY}) can be promoted (integrated) to a more restrictive
non-homogeneous equation with degree lower by one (i.e. one of the free parameters
in its solution can be fixed):
\be
\prod_{ \stackrel{Q\neq Q_{_0}}{Q\in R^{\otimes m}}}
{\hat\nabla^{(m)}_{\!q^{2 (\varkappa_{Q_{_0}} -\varkappa_Q)}\,}} {^F\!H_R^{[m,n]}} =
C_{Q_{_0}}(q,A)\cdot D_{Q_{_0}}(q,A)\cdot
\prod_{ \stackrel{Q\neq Q_{_0}}{Q\in R^{\otimes m}}}
\Big(1-q^{ 2 (\varkappa_{Q_{_0}} -\varkappa_Q)} \Big)
\label{nheqHOMFLY}
\ee
These framings are usually different from   topological
\be
{\cal F}_{\!R}^{\,n}
= \left(A^{|R|}q^{-4 
\varkappa_R}\right)^{n(m-1)}
\ee
thus one needs additional care
when considering Reidemeister-equivalent knot diagrams,
like $[m,n]$ and $[m,-n]$ in the case of torus knots.
{\it Here and in what follows we denote polynomials in topological framing
and associated quantities by calligraphic letters.}

\bigskip

In this paper we

(a) consider only fundamental representation $R=[1]=\Box$ and

(b) make the choice $Q_0=mR=[m]$ with $\varkappa_{[m]}=\frac{m(m-1)}{2}$,
and omit indices $f$ in what follows.

\noindent
Deviation from topological invariance in this case implies that
\be
H^{[m,-n]}_{_\Box}(A,q) = \left(\frac{{\cal F}_{_\Box}}{F_{_\Box}}\right)^{2n}
 \cdot H^{[m,n]}_{_\Box}(A^{-1},q)
= \left(\frac{{\cal F}_{_\Box}}{F_{_\Box}}\right)^{2n}\cdot H^{[m,n]}_{_\Box}(A^{-1},q^{-1})
\label{Hinvertedn}
\ee
and
\be
^F\! H_{_\Box} - \frac{F_{_\Box}}{{\cal F}_{_\Box}} \ \ \vdots \ \ \{Aq\}\{A/q\}
\ee
The second equality in (\ref{Hinvertedn}) follows from
transformation law for HOMFLY polynomial
from $R=\Box$ to transposed $R^{tr} = \Box = R$
and thus is an accidental feature of the fundamental representation.
Relation (\ref{Hinvertedn})
persists  in the case of superpolynomials,
but generically it does not survive in the case of their KR reductions.

\bigskip

At $m=2$ the fundamental HOMFLY in this framing is
\be
H^{[2,n]}_{_\Box} = \frac{\{Aq\}\{A\}}{\{q\}\{q^2\}} -
q^{2n}\cdot \frac{\{A\}\{A/q\}}{\{q\}\{q^2\}}
\ee
and satisfies
\be
\hat\nabla^{(2)}_{\!q^4\,}  H^{[2,n]}_{_\Box} = \frac{\{Aq\}\{A\}}{\{q\}\{q^2\}}\cdot  (1-q^4)
\ee
Associated Jones polynomial (i.e. HOMFLY at $A=q^2$)
\be
J^{[2,n]}_{_\Box} = [3] -q^{2n}
\ee
satisfies the same
\be
\hat\nabla^{(2)}_{\!q^4\,}  J^{[2,n]}_{_\Box} = [3]\cdot(1-q^4)
\ee
Here and below we use the standard notation:
$\{x\}=x-x^{-1}$ and $[k] = \frac{\{q^x\}}{\{q\}}$.

\bigskip

At $m=3$
\be
H^{[3,n]}_{_\Box} = \frac{\{Aq^2\}\{Aq\}\{A\}}{\{q\}\{q^2\}\{q^3\}}
- q^{2n}\cdot \frac{\{Aq\}\{A\}\{A/q\}}{\{q\}^2\{q^3\}}
+q^{4n}\cdot \frac{\{A\}\{A/q\}\{A/q^2\}}{\{q\}\{q^2\}\{q^3\}}
\ee
satisfies the second-order difference equation
\be
\hat\nabla^{(3)}_{\!q^6\,}\hat\nabla^{(3)}_{\!q^{12}\,}  H^{[3,n]}_{_\Box} =
(1-q^6)(1-q^{12})\cdot \frac{\{Aq^2\}\{Aq\}\{A\}}{\{q\}\{q^2\}\{q^3\}}
\ee
while associated Jones polynomial
\be
J^{[3,n]}_{_\Box} = [4] -[2]\cdot q^{2n}
\ee
satisfies just the first-order one:
\be
\hat\nabla^{(3)}_{\!q^6\,}  J^{[3,n]}_{_\Box} = [4]\cdot(1-q^6)
\ee

\bigskip

At $m=4$
\be
H^{[4,n]}_{_\Box} = \frac{\{Aq^3\}\{Aq^2\}\{Aq\}\{A\}}{\{q\}\{q^2\}\{q^3\}\{q^4\}}
- q^{2n}\cdot \frac{\{Aq^2\}\{Aq\}\{A\}\{A/q\}}{\{q\}^2\{q^2\}\{q^4\}}
+ 0\cdot q^{3n}\cdot \frac{ \{Aq\}\{A\}^2\{A/q\}}{\{q\}\{q^2\}^2\{q^3\}}
+ \nn\\
+\,q^{4n}\cdot \frac{\{Aq\}\{A\}\{A/q\}\{A/q^2\}}{\{q\}^2\{q^2\}\{q^4\}}
- q^{6n}\cdot \frac{ \{A\}\{A/q\}\{A/q^2\}\{A/q^3\}}{\{q\}\{q^2\}\{q^3\}\{q^4\}}
\ee
we get the first   degeneracy: $C_{[2,2]}=0$.
Thus the equation is of the order three,
$\#_{_{\rm Young\ diagrams}} - 1 - \#_{_{C=0}} =5-1-1=3$:
\be
\hat\nabla^{(4)}_{q^8}\hat\nabla^{(4)}_{\!q^{16}}\hat\nabla^{(4)}_{\!q^{24}\,}  H^{[4,n]}_{_\Box} =
(1-q^8)(1-q^{16})(1-q^{24})\cdot \frac{\{Aq^3\}\{Aq^2\}\{Aq\}\{A\}}{\{q\}\{q^2\}\{q^3\}\{q^4\}}
\ee
This time the degree of equation is further decreased at $N=3$ and $N=2$:
\be
H^{[4,n]}_{_\Box}(A=q^3,q) = \frac{[5][6]}{[2]} - [5][3]\cdot q^{2n} + [3]\cdot q^{4n}
\ee
satisfies the second-order
\be
\hat\nabla_{\!q^8\,} \hat\nabla_{\!q^{16}\,} H^{[4,n]}_{_\Box}(A=q^3,q)
= \frac{[5][6]}{[2]}\cdot(1-q^8)(1-q^{16})
\ee
while Jones polynomial
\be
J^{[4,n]}_{_\Box} = H^{[4,n]}_{_\Box}(A=q^2,q) = [5]-[3]\cdot q^{2n}
\ee
satisfies the first-order
\be
\hat\nabla_{\!q^8\,}  J^{[4,n]}_{_\Box} = [5]\cdot(1-q^8)
\ee

\bigskip

With increasing strand number $m$ we get the following pattern:

\bigskip

\be
\arraycolsep=0.5mm
{\footnotesize
\begin{array}{c||cccc||c||cccc}
m&& q^N=A: &\hspace{-0.5cm} {\rm HOMFLY}  &\hspace{-1cm}\rightarrow{\rm super} & \ldots
& N=2: \!\!\!\!\!\!\!\! &\!\!\!{\rm Jones} 
&\!\!\!\!\rightarrow{\rm Khovanov} \\
&&&&&&&\\
\hline
&&&&&&&&& \\
&\#_{_{\rm Young\ diagrams}} & \#_{_{C=0}} & 
\#_{\tiny\begin{array}{p{2cm}}pairs of \\coincident\\eigenvalues\end{array}}
&\ \stackrel{\rm recursion}{_{\rm degree}}
& & \#_{_{\rm Young\ diagrams}} & \#_{_{C=0}} &
\#_{\tiny\begin{array}{p{2cm}}pairs of \\coincident\\eigenvalues\end{array}}
&\ \stackrel{\rm recursion}{_{\rm degree}} \\
&&&&&&&&& \\
\hline
&&&&&&&&\\
2 &2&0&0&1&&2&0&0&1\\
3 &3&0&0&2&&2&0&0&1 \\
4 &5&1\rightarrow 0&0&3\rightarrow 4&&3&1\rightarrow 0&0&1\rightarrow 2 \\
5 &7&2\rightarrow 0&0&4\to 6&&3&1\rightarrow 0&0&1\rightarrow 2 \\
6 &11&5\rightarrow 0&2&6\to 9&  &4&2\rightarrow 0&1&1\rightarrow 2 \\
&&&&&&&\\
\ldots &&\ldots&&&\ldots&&\ldots&\\
&&&&&&&\\
m &{\rm coeff}_{q^m}\!\!\left(\frac{1}{\prod(1-q^{k})}\right)&&&&&
{\rm entier}\!\left(\frac{m+2}{2}\right)&& \\
&&&&&&&
\end{array}
}
\label{HOMFLYredtable}
\ee

\noindent
Arrows mark the changes which take place at the level of super-
and Khovanov polynomials.

\bigskip

Reversing the logic,
evolution recursion describes generic solutions of the
homogeneous difference equation: from
\be
\hat\nabla^{(m)}_{x_1}\ldots \hat\nabla^{(m)}_{x_p} H_n = 0
\label{homoeq}
\ee
it follows that
\be
H_n = \sum_{i=1}^p C_i\cdot x_i^{n/m}
\ee
with $p$ arbitrary integration constants $C_i$.
Each of these parameters appears in non-homogeneous equation
of degree $p-1$, obtained by integrating (\ref{homoeq})
w.r.t. one of the commuting difference operators $\hat\Delta_{x_k}$:
\be
\left(\prod_{i\neq k}^p \hat\nabla^{(m)}_{x_i} \right)\! H_n
= C_k \cdot x_k^{n/m} \cdot \prod_{i\neq k}^p \left(1-\frac{x_i}{x_k} \right)
\ee
or
\be
\left(\prod_{i\neq k}^p \hat\nabla^{(m)}_{x_i/x_k} \right)\! (x_k^{-n/m} H_n)
= C_k  \cdot \prod_{i\neq k}^p \left(1-\frac{x_i}{x_k} \right)
\ee
where the r.h.s. is independent of $n$.

Alternatively parameters $C_k$ can be defined from "initial conditions",
i.e. the values of $H_n$ at some $p$ different values of the evolution parameter
("time") $n$.
Used for this purpose below are the directly calculated KR polynomials
for relatively small knots.
Things would be greatly simplified if also
their twins at negative values of $n$ could be used.
However, the transformation $n\leftrightarrow -n$ acts badly
on the evolution formulas for KR polynomials,
and this way to extend the set of available
initial conditions does not seem to work.

\section{Recursion for torus superpolynomials}

Superpolynomials depending on one extra deformation parameter
${\bf t}\neq -1$  could be defined
from an analogue of the rule (\ref{AdamsHOMFLY})
with MacDonald instead of the Schur polynomials
-- and already this would make them depending on ${\bf q}$ and ${\bf t}$, --
but actually there are additional
correcting $c$-factors, discovered in \cite{DMMSS} and better described in two different
ways in \cite{Che} and \cite{HallLittl}.

Explicit expressions for the simplest {\it reduced} superpolynomials in topological framing
are \cite{DMMSS}:

{\footnotesize
$$
{\cal P}_{_\Box}^{[2,n]} = \left(\frac{A\cdot q}{t}\right)^n
\cdot \frac{ \{Aq\} \cdot q^{-n} -  \{A/t\}\cdot t^n}{\{qt\} }
$$

$$
{\cal P}_{_\Box}^{[3,n]} = \left(\frac{A\cdot q}{t}\right)^{2n}
\cdot \left(\frac{\{Aq^2\}\{Aq\}  \cdot q^{-2n}}{\{q^2t\}\{qt\} }
- (t^{\pm 2}+1+q^{\mp 2})\cdot
\frac{\{Aq\} \{A/t\}\cdot \left(t/q\right)^{2(n\mp 1)/3}}{\{q^2t\}\{qt^2\} }
+ \frac{ \{A/t\}\{A/t^2\}\cdot t^{2n} }{\{qt^2\}\{qt\} }
\right)
$$
}

\vspace{-1.5cm}

\be  \label{supps} \ee
{\footnotesize
$$
{\cal P}_{_\Box}^{[4,n]} = \left(\frac{A\cdot q}{t}\right)^{3n}
\cdot \left(\frac{\{Aq^3\}\{Aq^2\}\{Aq\} \cdot q^{-3n} }{\{q^3t\}\{q^2t\}\{qt\} }
+  \frac{\{t/q\}\cdot\{Aq\} \{Aq/t\}\{A/t\}\cdot (t/q)^n}{\{q^2t\}\{qt^2\}\{qt\}^2 }
- \frac{ \{A/t\}\{A/t^2\}\{A/t^3\}\cdot t^{3n}}{\{qt^3\}\{qt^2\}\{qt\} } -
\right.
$$
$$
\left\{  \begin{array}{ccc}
\left.
- \frac{(qt^2+q+q^{-1}+q^{-3})\cdot\{Aq^2\}\{Aq\} \{A/t\}\cdot q^{-n}\cdot(t/q)^{(n-1)/2}}
{\{q^3t\}\{q^2t^2\}\{qt\}  }
+ \frac{(t^4q^{-1}+t^2q^{-1}+q^{-1}+q^{-3} )\cdot \{Aq\} \{A/t\}\{A/t^2\}
\cdot t^{n}\cdot(t/q)^{(n-3)/2}}
{\{q^2t^2\}\{qt^3\}\{qt\} }
\right)&&n=1 \ {\rm mod}\ m \\
\left.
- \frac{(q^4t^{-1}+q^2t^{-1}+t^{-1}+t^{-3})\cdot\{Aq^2\}\{Aq\} \{A/t\}
\cdot q^{-n}\cdot(t/q)^{(n+3)/2}}
{\{q^3t\}\{q^2t^2\}\{qt\}  }
+ \frac{(tq^2+t+t^{-1}+t^{-3} )\cdot \{Aq\} \{A/t\}\{A/t^2\}
\cdot t^{n}\cdot(t/q)^{(n+1)/2}}
{\{q^2t^2\}\{qt^3\}\{qt\} }
\right) && n=-1 \ {\rm mod}\ m
\end{array} \right.
$$

$$
\ldots
$$
}

\noindent
where $n =\pm 1 \ {\rm mod} \ m$.
Note that  this parameter
takes the opposite values for the knots $[m,-n]$ and $[m,n]$ in (\ref{invn}).
Here we used our usual notation $\{x\} = x-x^{-1}$ and
the standard MacDonald variables $A,q,t$, used in \cite{DMMSS},
are related to DGR variables of \cite{DGR}
by $A^2 = -{\bf a^2t}$, $q=-{\bf qt}$, $t={\bf q}$.
Torus superpolynomials can be converted from Laurent to
true superpolynomials by extracting a power of $A/t=\sqrt{-{\bf a^2t}/{\bf q^2}}
\longrightarrow \sqrt{-{\bf q}^{2N-2}{\bf t}}$: \
${\cal P}^{[m,n]}_{[1]} = (A/t)^{(m-1)(n-1)}\cdot P^{[m,n]}_{[1]}
= (A/t)^{(m-1)(n-1)}\cdot\Big(1+ O(A^2,t,q)\Big)$ --
note that
$w=(n-1)(m-1)$ appears here instead of the writhe number $n(m-1)$
in the exponents of the framing factor in (\ref{supps}).
This explains the appearance of normalization factors in the following sections
of this paper.

\bigskip {\footnotesize
Superpolynomials are supposed to be {\it positive} Laurent polynomials
of the DGR variables ${\bf a}, {\bf q}, {\bf t}$,
i.e., all the coefficients should be non-negative integers.
However, this is not fully true for above expressions:
with the change of $n$ they switch from pure positive to pure negative
polynomials.
Moreover, this is a typical {\it anomaly}:
by insertion of additional overall factor of $(-)^{\frac{n-1}{m}}$
one could cure the problem for all positive values of $n$, but then all the
polynomials with negative $n$ will be pure negative --
violating also the invariance (\ref{invn}).
Sometime this anomaly can cause serious problems for construction of
superpolynomials \cite{Sat}, but for the torus knots {\it per se}
this is a rather innocent detail, which, however, should be remembered
and taken into account.
It leads to a minor modification in the case of {\it reduced}
super- and KR polynomials, while for the {\it unreduced} ones
it is also minor, but can look somewhat unfamiliar, see below.
}

Expressions (\ref{supps}) for the superpolynomials have the form
\be
\arraycolsep=0.5mm
\begin{array}{|c|c||l|ll|l|l|}
\cline{1-2}
m&n&\multicolumn{1}{c}{\mathcal{P}^{[n,m]}_{\Box}}\\
\cline{1-5}
\rule{0em}{1.2em}
2&2k\!+\!p&V_{r|2}\left(A^2/t^2\right)^n&+V_{r|11}\left(A^2q^2\right)^n&\\
\cline{1-6}
\rule{0em}{1.2em}
3&3k\!+\!p&V_{r|3}\left(A^6/t^6\right)^n&+V_{r|21}\left(A^6\right)^n&&+V_{r|111}\left(A^6q^6\right)^n\\
\hline
\rule{0em}{1.2em}
4&4k\!+\!p&V_{r|4}\left(A^{12}t^{-12}\right)^n&+V_{r|31}\left(A^{12}q^{6}t^{-10}\right)^n&
+V_{r|22}\left(A^{12}q^{8}t^{-8}\right)^{n}&
+V_{r|211}\left(A^{12}q^{10}t^{-6}\right)^n&+V_{r|1111}\left(A^{12}q^{12}\right)^n\\
\hline
\multicolumn{2}{|c||}{N\ge}&\multicolumn{1}{c|}{1}&\multicolumn{2}{c|}{2}
&\multicolumn{1}{c|}{3}&\multicolumn{1}{c|}{4}\\
\hline
\end{array}\!\!\!\!\!\!\!
\label{spev}
\ee
The last line in this table is to remind  that for $N\le m$ only
the eigenvalues from the first $N$ columns do contribute to the answer,
while the coefficients  $V$ of the remaining ones vanish in these cases.
Such polynomials satisfy the recursion relations
\be
\arraycolsep=0.5mm
\begin{array}{lllllllr}
&&&
\nabla^{(2)}_{A^2q^2}&
\nabla^{(2)}_{A^2/t^2}&
P^{[2,n]}_{\square}=0\\
\rule{0em}{1.5em}
&&\nabla^{(3)}_{A^6q^6}
&\nabla^{(3)}_{A^6}
&\nabla^{(3)}_{A^6/t^6}
&P^{[3,n]}_{\square}=0\\
\rule{0em}{1.5em}
\nabla^{(4)}_{A^{12}q^{12}}&
\nabla^{(4)}_{A^{12}q^{10}t^{-6}}&
\nabla^{(4)}_{A^{12}q^{8}t^{-8}}&
\nabla^{(4)}_{A^{12}q^{6}t^{-10}}&
\nabla^{(4)}_{A^{12}t^{-12}}&
P^{[4,n]}_{\square}=0\\
&&&\multicolumn{2}{r}{\underbrace{\hspace{8em}}_{N\ge 2}}\\[-1em]
&\multicolumn{4}{r}{\underbrace{\hspace{17em}}_{N\ge 3}}\\[-1em]
\multicolumn{5}{r}{\underbrace{\hspace{21em}}_{N\ge 4}}
\end{array}
\label{tfrdiff}
\ee
generated by the difference operators
\be
\hat\nabla^{(m)}_{\lambda}P^{m,n}_{\square}\equiv
P^{m,n}_{\square} -\lambda\cdot P^{m,n-m}_{\square}.
\label{nabdef}
\ee
Since for $N\le m$ the polynomials $P_{\square}^{n,m}$ in fact depend only on the first few eigenvalues, they actually satisfy  the shortened recursions, which are generated by the difference operators, which survive in  the last line in the table (\ref{tfrdiff}).

Equations (\ref{tfrdiff}) can be integrated to
\be
\hat\nabla^{(2)}_{q^2t^2} \Big((A/t)^{-n}\cdot{\cal P}^{[2,n]}_{_\Box}\Big) = const_n \nn \\
\hat\nabla^{(3)}_{q^4t^2} \Big((A/t)^{-2n}\cdot{\cal P}^{[3,n]}_{_\Box}\Big) = const_n
\ {\rm mod} \ \{A/t^2\}
\label{shortrecursion} \\
\hat\nabla^{(4)}_{q^6t^2} \hat\nabla^{(4)}_{q^8t^4}
\Big((A/t)^{-3n}\cdot{\cal P}^{[4,n]}_{_\Box}\Big) = const_n
\ {\rm mod} \ \{A/t^2\}, \nn  \\
\ldots
\nn\ee
if we multiply each $\mathcal{P}_{\Box}$ on the normalisation factor that makes the eigenvalue for $Q_0=[m]$ to be equal $1$.
Recursions for $m=3$ and $m=4$ here are the short ones, true for $A=t^2$, when $\{A/t^2\}=0$,
i.e., they are valid for (super)Jones polynomials at $N=2$:
for generic $N$ the degrees of the equations would be $3$ and $5$
(or $4$ for HOMFLY in the case of $m=4$, when representation $[22]$ does not contribute).
Note that for  $m=4$ one does {\it not} expect difference equations
with $\hat\nabla^{(m/2)}$,
because the coefficients in the superpolynomials depend on the residue $n\, {\rm mod}\, m$.

Lyapunov exponents for the torus $n$-evolution are made from MacDonald split of
the Casimir eigenvalues: for the Young diagram $Q=\{Q_1\geq Q_2\geq \ldots \geq Q_l>0\}$
\be
\lambda_Q^{n} = \left(\frac{Aq}{t}\right)^{n(m-1)}
\cdot (q^{-\nu(Q^{tr})}\cdot t^{\nu(Q)})^{2n/m}
= \left(-{\bf a^2t}\right)^{n(m-1)}\cdot
{\bf q}^{-\frac{2n\varkappa(Q)}{m}}\cdot (-{\bf t})^{-\frac{2n\nu(Q^{tr})}{m}},
\ee
where $\nu(Q) = \sum_{i=1}^l (i-1)Q_i$ and
$\nu(Q)-\nu(Q^{tr}) =\varkappa(Q) = \sum_{(i,j)\in Q} (i-j)$.
The spectra of the $m$-th powers  $\left(\frac{\lambda_Q}{\lambda_{[m]}}\right)^m$,
which enter the difference operators in the {\it short} recursion and are independent
of $A$ and, consequently, on the details of the $N$-reduction,
can be extracted from \cite{DMMSS}:

%
%

\be
{\footnotesize
\arraycolsep=0.5mm
\hspace{0cm}
\begin{array}{|c|c|ccc|cccc|ccc|cc|c|c|c|}
\cline{1-5}
\rule{0mm}{3.5mm}
2 &[2]&[11] &&&\multicolumn{12}{|c}{}\\[1mm]
&1 &{\bf q^4t^2} &&&\multicolumn{12}{|c}{}\\
\cline{1-9}
\rule{0mm}{4mm}
3 &[3]&[21]&&&\left[1^3\right]&&&&\multicolumn{8}{|c}{}\\[1mm]
& 1 &{\bf q^6t^4} &&&{\bf q^{12}t^6} &&&&\multicolumn{8}{|c}{}\\
\cline{1-12}
\rule{0mm}{4mm}
4 & [4]&[31]&[22]&&\left[21^3\right]&&&&\left[1^4\right]&&&\multicolumn{5}{|c}{}\\[1mm]
&1 &{\bf q^8t^6}&{\bf q^{12}t^8} &&{\bf q^{16}t^{10}} &&&&{\bf q^{24}t^{12}}&&&\multicolumn{5}{|c}{}\\
\cline{1-14}
\rule{0mm}{4mm}
5 &[5]&[41]&[32]&&[311]&[221]&&&\left[21^3\right]&&&\left[1^5\right]&&
\multicolumn{3}{|c}{}\\[1mm]
& 1 &  {\bf q^{10}t^8}&{\bf q^{16}t^{12}}&&{\bf q^{20}t^{14}}&{\bf q^{24}t^{16}}&&
&{\bf q^{30}t^{18}}&&&{\bf q^{40}t^{20}}&&
\multicolumn{3}{|c}{}\\
\cline{1-15}
\rule{0mm}{4mm}
6&[6]&[52]&[42]&[33]&[411]&[321]&[222]&&\left[31^3\right]&[2211]&&\left[21^4\right]&&\left[1^6\right]&
\multicolumn{2}{c}{}\\[1mm]
& 1 & {\bf q^{12}t^{10}}&{\bf q^{20}t^{16}}&{\bf q^{24}t^{18}}
& {\bf q^{24}t^{18}}&{\bf q^{30}t^{22}}&{\bf q^{36}t^{24}}&
&{\bf q^{36}t^{24}}&{\bf q^{40}t^{26}}&&{\bf q^{48}t^{28}}&&{\bf q^{60}t^{30}}&\multicolumn{2}{c}{}\\
\cline{1-16}
\rule{0mm}{4mm}
7  &[7]&[61]&[52]&[43]&[511]&[421]&[331]&[322]&[4111]&[3211]&[2221]&
\left[31^4\right]&\left[221^3\right]&\left[21^6\right]&\left[1^7\right]\\[1mm]
&1 &  {\bf q^{14}t^{12}} 
& {\bf q^{24}t^{20}}& {\bf q^{30}t^{24}}
& {\bf q^{28}t^{22}}& {\bf q^{36}t^{28}}& {\bf q^{40}t^{30}}& {\bf q^{44}t^{32}}
& {\bf q^{42}t^{30}}& {\bf q^{48}t^{34}}& {\bf q^{54}t^{36}}
& {\bf q^{36}t^{56}} 
& {\bf q^{60}t^{38}}&  {\bf q^{70}t^{40}}&{\bf q^{84}t^{42}}\\
\hline
\ldots&\ldots&\multicolumn{3}{|c|}{\ldots}&\multicolumn{4}{|c|}{\ldots}
&\multicolumn{3}{|c|}{\ldots}&
\multicolumn{2}{|c|}{\ldots}&\ldots&\ldots&\ldots\\
\hline
\rule{0mm}{4mm}
N\!\!\ge&1&\multicolumn{3}{|c|}{2}&\multicolumn{4}{|c|}{3}&\multicolumn{3}{|c|}{4}&
\multicolumn{2}{|c|}{5}&6&7&\ldots \\
\hline
\end{array}
\label{evs}
}
\ee

KR polynomials
${K}^{\cal K}_R(N,{\bf q},{\bf t})$
are related to superpolynomial \cite{GSV}
by the DGR  rule \cite{DGR}:\footnote{
The standard definition in the literature actually includes
additional factor:
${K}_{R}^{\cal K}(N,{\bf q},{\bf t}) =
 (-{\bf t})^{-(m-1)(n-1)/2}\cdot
\left.
{P}_{R}^{\cal K}({\bf a},{\bf q},{\bf t}) \ {\rm mod} \
\{A/t^N\}
 \Big({\bf a^2 t} + {\bf q}^{2N}\Big)
\right|_{{\bf a}={\bf q}^N}$.
We, however, omit it because this leads to significant
simplifications in the logic, while it is always easy to
restore the factor in the answers.
Note  that, like the superpolynomials (\ref{supps})
our ${\cal K}^{[m,n]}$ can be pure negative rather than pure positive.
Also the {\it reduced} $N=2$ polynomials computed with the standard programs~\cite{katlas, gorpols}
contain one more extra factor ${\bf q}^{-1}$,
which we also omit.
}
\be
 {K}_{R}^{\cal K}(N,{\bf q},{\bf t}) =
\left.
{P}_{R}^{\cal K}({\bf a},{\bf q},{\bf t}) \ {\rm mod} \
\{A/t^N\}
\right|_{{\bf a}={\bf q}^N}
\label{KRfroSUP}
\ee
The choice of the coefficient in front of the "differential"
\be
d_N = 1+\frac{{\bf a^2 t}}{{\bf q}^{2N}} \ \stackrel{{\bf a}={\bf q}^N}{\longrightarrow} \
1 + {\bf t}
\label{diffN}
\ee
is, however, not specified.
In other words,
\be
K_R^{\cal K}(N,{\bf q},{\bf t}) =
P_R^{\cal K}({\bf a}={\bf q}^N,{\bf q},{\bf t})\ \ -
\underbrace{(1+{\bf t})}_{d_N({\bf a}={\bf q}^N,{\bf q},{\bf t})}
\!\!\cdot\ \ \xi_R^{\cal K}(N,{\bf q},{\bf t})
\label{xidef}
\ee
and additional principles are needed to separate positive $\xi$ and positive ${\cal K}$,
which also contains positive terms divisible by $1+{\bf t}$ --
which {\it could be}, but are {\it not} moved to $\xi$.
The search for these principles is a big challenge, but it is beyond the scope of
the present text.


\section{Recursion for torus Khovanov polynomials at $m=2$}

In this case
Khovanov polynomials are known from direct calculation at all odd $n$
\cite{Kh,BN, DGR, GGS,DM}
and they actually coincide
with the result of the substitution of $N=2$ into
\be
(-{\bf q^{2(N-1)}t})^{-k}\cdot
{\cal P}^{[2,2k+1]}_{r} \ = \ \frac{A}{t}\cdot \left(
\frac{M^*_{[2]}}{M^*_{[1]}}
-  ({\bf q^2t})^{2k+1}\cdot\frac{\{{\bf q}^2\}}{\{{\bf q^2t}\}}
\cdot \frac{M^*_{[11]}}{M^*_{[1]}}\right) = \nn
\ee
\vspace{-0.5cm}
\be
=\ \frac{1-({\bf q^2t})^{2k+2}}{1-({\bf q^2t})^2}
+ {\bf q}^{2N+2}{\bf t}^3\cdot\frac{1-({\bf q^2t})^{2k}}{1-({\bf q^2t})^2}
\ \ = \ \  1 + {\bf q^4 t^2}\Big(1+{\bf q}^{2N-2}{\bf t}\Big)
\cdot\frac{1-({\bf q^2t})^{2k}}{1-({\bf q^2t})^2}
\label{Pr2str}
\ee
for {\it reduced} and
\be
(-{\bf q^{2(N-1)}t})^{-k}\cdot
{\cal P}_{ur}^{[2,2k+1]} \ = \
[N] + {\bf q^3t^2}(1+{\bf q}^{2N}{\bf t})\cdot
\frac{1-({\bf q^2t})^{2k}}{1-({\bf q^2t})^2}\cdot[N-1]
\label{Pur2str}
\ee
for {\it unreduced} superpolynomials.
As explained in \cite{AnoM}, relation between the two  is implied by
the differential expansion of the {\it reduced} fundamental superpotential:
\be
{\cal P}_r = 1 + {\cal F}\{Aq\}\{A/t\}
\ee
and for $N$-reduction  $\{A/t\} \longrightarrow (1+{\bf q^{2N-2}t})$.
Naively {\it unreduced} polynomial is obtained by multiplication
with the quantum number $[N]$.
However, one can use the usual identification rule \cite{AnoM}
\be
(1+{\bf q^2t})\cdot[k] \cong \frac{1}{{\bf q}^{k-1}}(1+{\bf q^{2k}t})
\label{redqn}
\ee
to substitute
\be
\begin{array}{cccccc}
&&&{\bf q}^{N-2}(1+{\bf q^2t})\cdot[N][N-1]\\
&&\swarrow&&\searrow\\
&(1+{\bf q^{2N-2}t})\cdot[N]\!\!\!\!\! &&\cong&&
\!\!\!\!\!\frac{1}{\bf q}\cdot(1+{\bf q^{2N}t})\cdot [N-1]\\
\end{array}
\label{reduct}
\ee
-- and this is how (\ref{Pur2str}) is obtained from (\ref{Pr2str}).

\bigskip

Putting $N=2$, we obtain from (\ref{Pr2str})
\be
{\cal K}^{[2,2k+1]}_r = \ \FF^{k}\cdot K^{[2,2k+1]}_r
= \FF^{k}\cdot
\left(1+ {\bf q^4t^2}\cdot\frac{1-({\bf q^2t})^{2k}}{1-{\bf q^2 t}}\right)
\label{Kr2}
\ee
for {\it reduced} case and
\be
{\cal K}^{[2,2k+1]}_{ur}  = \ \FF^{k}\cdot K^{[2,2k+1]}_{ur}
=  \FF^{k}\cdot\left({\bf q}+{\bf q}^{-1} + {\bf q^3t^2}(1+{\bf q^4\bf t})\cdot
\frac{1-({\bf q^2t})^{2k}}{1-({\bf q^2 t})^2}\right)
\label{Kur2}
\ee
from (\ref{Pur2str}) for the {\it unreduced} one.
Both of them satisfy non-homogeneous first-order equations:
\be
\nabla^{(2)}_{\bf q^4 t^2} K_r^{[2,n]} = K_r^{[2,n]}-{\bf q^4t^2}K_r^{[2,n-2]}
= 1+{\bf q^6t^3} \nn
\ee
\vspace{-0.6cm}
\be
\nabla^{(2)}_{\bf q^4 t^2} K_{ur}^{[2,n]}
= {\bf q}+{\bf q}^{-1}-{\bf q^5t^2}+{\bf q^7t^3}
= ({\bf q}+{\bf q}^{-1})(1+{\bf q^6t^3}) - \underline{{\bf q^5 t^2}(1+{\bf t})}
\label{eqsK2}
\ee
Note that these are exactly the same recursions as the one in the first line
of (\ref{shortrecursion}):  $q^2t^2 = {\bf q}^4{\bf t}^2$,
only now we provided explicit expressions for the right hand sides.
One can eliminate these right hand sides by applying one more difference operator,
then the equation is the same for {\it reduced} and {\it unreduced} polynomials:
\be
\nabla^{(2)}_{1}\nabla^{(2)}_{\bf q^4 t^2} {K}^{[2,n]}_{_\Box} =
K^{[2,n]}_{_\Box} -(1+{\bf q^4t^2})K^{[2,n-2]}_{_\Box} + {\bf q^4t^2}K^{[2,n-4]}_{_\Box}=0
\nn \ee
\vspace{-0.6cm}
$$\updownarrow $$
\vspace{-0.6cm}
\be
\nabla^{(2)}_{-{\bf q^2t}}\nabla^{(2)}_{-{\bf q^6 t^3}} {\cal K}^{[2,n]}_{_\Box} =
{\cal K}^{[2,n]}_{_\Box} + ({\bf q^2t}+{\bf q^6t^3}){\cal K}^{[2,n-2]}_{_\Box}
+ {\bf q^8t^4}{\cal K}^{[2,n-4]}_{_\Box}=0
\ee
with
\be
A^2/t^2=-{\bf a^2t/q^2}\ \stackrel{\ \ \ {\bf a}={\bf q}^2 }{\longrightarrow}\ -{\bf q^2t}
\nn\\
A^2q^2=-{\bf a^2t^3q^2}\ \stackrel{\ \ \ \ {\bf a}={\bf q}^2\ }{\longrightarrow}\ -{\bf q^6t^3}
\!\!\!\!
\ee

Invariance w.r.t. inversion of $n$ would state:
\be
{\cal K}_{r}^{[2,-n]}({\bf q},{\bf t}) =
{\cal K}_{r}^{[2,n]}\left(\frac{1}{\bf q},\frac{1}{\bf t}\right)
\ \ \ \Longrightarrow \ \ \
 K_{r}^{[2,-n]}({\bf q},{\bf t})
 = -{\bf q^2t} \cdot
{K}_{r}^{[2,n]}\left(\frac{1}{\bf q},\frac{1}{\bf t}\right)
\label{invn2}
\ee
It is easy to check that the first recursion  for $K_r$ in (\ref{eqsK2}) is indeed invariant
under this transformation,
i.e. (\ref{invn2}) is indeed true -- what is not a surprise, because $K_r$ is obtained
by a change of variables from the topologically invariant {\it reduced}
superpolynomial (\ref{Pr2str}).
However, this is not the case for the {\it unreduced} $K_{ur}$ --
like the {\it unreduced} superpolynomial (\ref{Pur2str}),
it changes non-trivially under the $n$-inversion (unless ${\bf t}=-1$),
because of the underlined term in (\ref{eqsK2}).
The second branch differs from (\ref{Kur2}) by a slight modification of the first term:
\be
\tilde K^{[2,2k+1]}_{ur}
= -{\bf qt}+{\bf q}^{-1} + {\bf q^3t^2}(1+{\bf q^4\bf t})\cdot
\frac{1-({\bf q^2t})^{2k}}{1-({\bf q^2 t})^2}
\label{Kur2t}
\ee
Instead of (\ref{invn2}) the evolution formula for the unreduced Khovanov polynomial
satisfies
\be
{\cal K}^{[2,-n]}_{ur}({\bf q},{\bf t}) =
 \tilde {\cal K}^{[2,n]}_{ur}\left(\frac{1}{\bf q},\frac{1}{\bf t}\right)
 \ \Longleftrightarrow \
K^{[2,-n]}_{ur}({\bf q},{\bf t}) =
-{\bf q^2t} \cdot \tilde K^{[2,n]}_{ur}\left(\frac{1}{\bf q},\frac{1}{\bf t}\right)
\label{invn2mod}
\ee

\bigskip

We can now inverse the logic:   {\it begin} from the recursion relation
(\ref{shortrecursion})
and {\it derive} the evolution formula for arbitrary $K^{[2,n]}_{_\Box}$.
Short and long recursions for $m=2$ mean that
\be
K^{[2,n]}_{_\Box} = \alpha'+\beta'\cdot \lambda^{2n} \ \Longleftrightarrow \
{\cal K}^{[2,n]} = \alpha \cdot\lambda^{n} + \beta\cdot\lambda^{3n}
\ee
with $\lambda=\FF^{1/2}$ and some $n$-independent $\alpha$ and $\beta$.
Note that at $N=2$ the two Lyapunov exponents differ by a factor of $3$,
which is independent of parameters ${\bf q}$ and ${\bf t}$.
To solve the recursion we need
{\it initial conditions} and the knowledge of just two
should be enough.
It can seem that we always have these two: $n=\pm 1$ correspond to the unknot.
The problem, however, is that we need (\ref{invn2}) to use the both.
If (\ref{invn2}) is true -- like it is for {\it reduced} KR polynomial, we immediately get:
\be
\left.\alpha \cdot\lambda^{n} + \beta\cdot \lambda^{3n}\right|_{n=\pm 1}
= {\cal K}^{unknot}_r = 1
\ \ \ \Longrightarrow  \nn
\ee
\vspace{-0.6cm}
\be
\Longrightarrow \ \ \
\boxed{
{_r{\cal K}}^{[2,n]}_{_\Box} = \FF^{^{\frac{n-1}{2}}}\cdot
\left(1+{\bf q^4 t^2}\cdot\frac{1-({\bf q^2t})^{n-1}}{1-{\bf q^2t}}\right)
=  \FF^{^{\frac{n+1}{2}}}\cdot
\left(1-\frac{1}{\bf q^2 t}\cdot\frac{1-({\bf q^2t})^{n+1}}{1-{\bf q^2t}}\right)
},
\label{KR2}
\ee
i.e., reproduce (\ref{Kr2}).
Note that the input as initial conditions was only the unknot (!) --
this illustrates the usual power of evolution method.
However, for {\it unreduced} KR polynomial we rather need  (\ref{invn2mod}), not (\ref{invn2}):
\be
\left.\alpha \cdot\lambda^n + \beta\cdot \lambda^{3n} \right|_{n=\pm 1}
= {\cal K}^{unknot_\pm}_{ur}
=  \left.{\bf q}+(-{\bf t})^{^{\frac{n-1}{2}}}\!\!\!\cdot{\bf q}^{-1}\right|_{n=\pm 1}
\ \ \ \Longrightarrow \nn \\ \nn \\
\Longrightarrow \ \ \
{_{ur}{\cal K}}^{[2,n]}_{_\Box} = \FF^{^{\frac{n-1}{2}}}\!\!\!\cdot
\left({\bf q}+{\bf q}^{-1}
+{\bf q^3 t^2}(1+{\bf q^4 t})\cdot\frac{1-({\bf q^2t})^{n-1}}{1-{\bf q^4t^2}}
\right)
\label{KUR2}
\ee
what coincides with (\ref{Kur2}).
For the dual branch we have
\be
\left.\tilde\alpha \cdot\lambda^n + \tilde \beta\cdot \lambda^{3n} \right|_{n=\pm 1}
= \tilde{\cal K}^{unknot_\pm}_{ur} =
\left. (-{\bf t})^{^{\frac{n+1}{2}}}\!\!\! \cdot{\bf q}+ {\bf q}^{-1} \right|_{n=\pm 1}
\ \ \ \Longrightarrow \nn \\ \nn \\
\Longrightarrow \ \ \
{_{ur}\tilde{\cal K}}^{[2,n]}_{_\Box}  =
\FF^{^{\frac{n-1}{2}}}\!\!\!\cdot
\left(-{\bf qt}+{\bf q}^{-1} + {\bf q^3t^2}(1+{\bf q^4\bf t})\cdot
\frac{1-({\bf q^2t})^{n-1}}{1-({\bf q^2 t})^2}\right) =
\nn
\ee
\vspace{-0.5cm}
\be
 = (\FF 
 )^{^{\frac{n+1}{2}}}\!\!\!\cdot
\left({\bf q}+{\bf q}^{-1}
- \frac{1+{\bf q^4 t}}{\bf q^3 t}\cdot\frac{1-({\bf q^2t})^{n+1}}{1-{\bf q^4t^2}}
\right)
\ \ \ \ \ \ \ \ \ \ \ \ \
\label{KUR2t}
\ee
${\cal K}^{[2,n]}_{ur}$ and $\tilde{\cal K}^{[2,n]}_{ur}$
are fully positive (or fully negative) respectively for $n>0$ and $n<0$,
while in the "foreign" domains they contain terms with different signs --
still they remain related by (\ref{invn2mod}).

There is also a direct way from (\ref{KR2}) to (\ref{KUR2}) and (\ref{KUR2t}):
instead of just multiplying by $[2]={\bf q}+{\bf q}^{-1}$ one should apply
the  rule (\ref{redqn}):
\be
\phantom.
[2]\cdot(1+{\bf q^2t})\longrightarrow \frac{1}{\bf q}\cdot (1+{\bf q^4t})
\label{reduct2}
\ee
Then we get from the first version of (\ref{KR2})
\be
\phantom.[2]\cdot\left(
1+{\bf q^4t^2}(1+{\bf q^2 t})\frac{1-({\bf q^2t})^{n-1}}{1-({\bf q^2t})^2}\right)
\ \longrightarrow \
{\bf q}+{\bf q}^{-1}+{\bf q^3t^2}(1+{\bf q^4 t})
\frac{1-({\bf q^2t})^{n-1}}{1-({\bf q^2t})^2}
\ee
i.e., exactly (\ref{KUR2}), while the second version turns into (\ref{KUR2t}).


\section{Recursion for Khovanov ($N=2$) polynomials at $m=3$ }

In this case recursion is almost as simple as at $m=2$.
For $N=2$ the r.h.s. with $\{A/t^2\}\stackrel{A=t^2}{=}0$ in (\ref{shortrecursion})
can be omitted,
{\it but} at the l.h.s. one should substitute ${\bf a}={\bf q}^2$,
i.e., $(A/t)^6=-{\bf q^{6}t^3}$:
this is a manifestation of the double-face reduction rule (\ref{KRfroSUP}).
Thus, {\it though this does not yet follow from any first principle argument},
one can {\it expect} that Khovanov polynomials at $N=2$
satisfy the {\it second} order homogeneous equation
\be
\hat\nabla^{(3)}_{-\bf q^6t^3} \hat\nabla^{(3)}_{-\bf q^{12}t^7}\,
 {\cal K}_{_\Box}^{[3,n]}(N=2) = 0
\label{heqs3}
\ee
while Khovanov--Rozansky at $N=3$ should satisfy the full-fledged third order one,
the same as the full superpolynomials with ${\bf a}={\bf q}^3$ and $(A/t)^6=-{\bf q^{12}t^3}$:
\be
\hat\nabla^{(3)}_{-{\bf q^{12}t^3} }
\hat\nabla^{(3)}_{-{\bf q^{18}t^7} } \hat\nabla^{(3)}_{-{\bf q^{24}t^9} } \,
{\cal K}_{_\Box}^{[3,n]}(N=3) = 0
\ee
We return to KR at $N=3$ in sec.\ref{N3} below, and concentrate in this section
on Khovanov polynomials with $N=2$.

\bigskip

Making use of explicit expressions~\cite{BN, gorpols}
%
%
we can check that this {\it short}-evolution {\it hypothesis} is indeed true
and restore entire evolution formulas:
\be
\boxed{
_r{\cal K}_{_\Box}^{[3,n]} =
\left\{
\begin{array}{ccc}\FF^{n-1}\cdot\left(1+
\Big(1+{\bf q^2t}+{\bf q^2t^2}+{\bf q^6t^3}\Big)\cdot {\bf q^4t^2}\cdot
\frac{1-({\bf q^6t^4})^{^{\frac{n-1}{3}}}}{1-{\bf q^6t^4}}\right) && n=1\ {\rm mod} \ 3
 \\
\FF^{n+1}\cdot (-{\bf t})^{-1}\cdot\left(1 -
\frac{1+{\bf q^4t}+{\bf q^4t^2}+{\bf q^6t^3}}{\bf q^4t}\cdot
\frac{1-({\bf q^6t^4})^{^{\frac{n+1}{3}}}}{1-{\bf q^6t^4}}\right) && n=-1\ {\rm mod} \ 3
\end{array}
\right.
}
\label{KR3}
\ee
With the help of (\ref{KR2}) we can check the
topological invariance under the change $[m,n]\longleftrightarrow [n,m]$:
\be
{\cal K}_r^{[3,2]}=  {\cal K}_r^{[2,3]}
\label{anor23}
\ee
However, there is a non-trivial factor
in the  mirror-symmetry relation:
\be
{\cal K}_r^{[3,-n]}({\bf q},{\bf t}) =
-\frac{1}{\bf t} \cdot {\cal K}_r^{[3,n]}\left(\frac{1}{\bf q},\frac{1}{\bf t}\right)
\label{invn3}
\ee
and, accordingly,
${\cal K}_r^{[3,-2]}=  -\frac{1}{\bf t}\cdot {\cal K}_r^{[3,2]} =
-\frac{1}{\bf t}\cdot {\cal K}_r^{[2,3]} = -\frac{1}{\bf t}\cdot {\cal K}_r^{[2,-3]}$.


\bigskip

Similarly, for {\it unreduced} Khovanov polynomials we get:
\be
\boxed{
_{ur}{\cal K}_{_\Box}^{[3,n]} =
\left\{
\begin{array}{ccc}\FF^{n-1}\cdot\left({\bf q}+{\bf q}^{-1}+
\big(1+ {\bf q^2t^2}+{\bf q^4t^2}\big)(1+{\bf q^4t})\cdot {\bf q^3t^2}\cdot
\frac{1-({\bf q^6t^4})^{^{\frac{n-1}{3}}}}{1-{\bf q^6t^4}}\right) && n=1\ {\rm mod} \ 3
 \\
\FF^{n+1}\cdot (-{\bf t})^{-1}\cdot\left({\bf q}+{\bf q}^{-1} -
\frac{(1+{\bf q^2}+{\bf q^4t^2})(1+{\bf q^4t })}{\bf q^5t}\cdot
\frac{1-({\bf q^6t^4})^{^{\frac{n+1}{3}}}}{1-{\bf q^6t^4}}\right) && n=-1\ {\rm mod} \ 3
\end{array}
\right.
}
\nn
\ee
{\footnotesize
Again, these formulas can be obtained directly from (\ref{KR3}) by application of
the rule (\ref{reduct2}), e.g.,
\be
\phantom. [2]\cdot\left( 1+
\Big((1+{\bf q^2t})+{\bf q^2t^2}(1+{\bf q^4t})\Big)\cdot {\bf q^4t^2}\cdot
\frac{1-({\bf q^6t^4})^{^{\frac{n-1}{3}}}}{1-{\bf q^6t^4}}\right)
\ \longrightarrow \
{\bf q} + {\bf q}^{-1} +
(1+{\bf q^4t})\Big(1 +[2]{\bf q^3t^2} \Big) \cdot {\bf q^4t^2}\cdot
\frac{1-({\bf q^6t^4})^{^{\frac{n-1}{3}}}}{1-{\bf q^6t^4}}
\nn
\ee
}
{\it Unreduced} polynomials satisfy topological identity
\be
{\cal K}_{ur}^{[3,2]}=  {\cal K}_{ur}^{[2,3]}
\label{Anor23}
\ee
and mirror relation
\be
{\cal K}_{ur}^{[3,-n]}({\bf q},{\bf t}) =
-\frac{1}{\bf t}\cdot {\cal K}_{ur}^{[3,n]}\left(\frac{1}{\bf q},\frac{1}{\bf t}\right)
\label{Invn3}
\ee

\section{Recursion for Khovanov ($N=2$) polynomials at $m=4$}

Proceeding to four strands, $m=4$, we can expect the third-order difference equation
for Khovanov polynomials,
\be
\hat\nabla^{(4)}_{\bf q^{12}t^6} \hat\nabla^{(4)}_{\bf q^{20}t^{12}}
\hat\nabla^{(4)}_{\bf q^{24}t^{14}}\,
 {\cal K}_{_\Box}^{[4,n]}(N=2) = 0
\label{heqs4}
\ee
This means that they should be linear combinations of three
Lyapunov exponentials -- and indeed they are.

\bigskip

\noindent
For $n=1 \, {\rm mod}\,4$
\be
{\cal K}_{r}^{[4,n]} = \FF^{^{\frac{3(n-1)}{2}}}\cdot\left(1-
\underline{(1+{\bf t})\cdot{\bf q^{10}t^5}(1+{\bf q^2t})
\cdot\frac{1-({\bf q^{12}t^8})^{^{\frac{n-1}{4}}}}{(1-{\bf q^4t^2})(1-{\bf q^6t^4})}}
\ +\right.
\nn \\
\left.+ {\bf q^4t^2}(1+{\bf q^6t^3})
\left(1+ \frac{{\bf q^2t}(1+{\bf t}+{\bf q^2t^3}+{\bf q^6t^4})}{1-{\bf q^4t^2}}\right)
\cdot\frac{1-({\bf q^8t^6})^{^{\frac{n-1}{4}}}}{ 1-{\bf q^8t^6} }
\right)
\ee
\be
{\cal K}_{ur}^{[4,n]} = \FF^{^{\frac{3(n-1)}{2}}}\cdot\left({\bf q}+{\bf q}^{-1}
-\underline{(1+{\bf t})\cdot{\bf q^9t^5}(1+{\bf q^4t})
\cdot\frac{1-({\bf q^{12}t^8})^{^{\frac{n-1}{4}}}}{(1-{\bf q^4t^2})(1-{\bf q^6t^4})}}  +
\right. \nn \\
\left.
+ \left({\bf q^3t^2}(1+{\bf q^2t^2})(1+{\bf q^4t^2})
+ \frac{{\bf q^7t^5}({\bf t}+{\bf q^2}+{\bf q^2t}+{\bf q^8t^4})}
{1-{\bf q^4t^2}} \right)\cdot(1+{\bf q^4t})
\cdot\frac{1-({\bf q^8t^6})^{^{\frac{n-1}{4}}}}{ 1-{\bf q^8t^6} }
\right)
\ee
Underlined structures
enter with the coefficient $(1+{\bf t})$, because it does not
contribute in the case of HOMFLY polynomial (the coefficient in front of $D_{[22]}$
is ``accidentally'' vanishing when $t=q$).
Despite it inters with negative sign, it does not spoil the positivity of
the entire expression at $n>0$.
However, at $n<0$ the polynomial fail to be positive or negative and consists of
monomials with different signs.
Thus there are no chances for any relation like (\ref{invn}).

\bigskip

For $n=-1 \, {\rm mod}\,4$
\be
{\cal K}_{r}^{[4,n]} = \FF^{^{\frac{3(n\blue{+}1)}{2}}}\cdot {\bf t}^{-2}
\cdot\left(-{\bf t}
+\underline{(1+{\bf t})\cdot\frac{1+{\bf q^2t}}{\bf q^2}
\cdot\frac{1-({\bf q^{12}t^8})^{^{\frac{n\blue{+}1}{4}}}}{(1-{\bf q^4t^2})(1-{\bf q^6t^4})}}\
- \right. \nn \\ \left.
- \frac{1+{\bf q^6t^3}}{\bf q^6t}
\cdot\frac{1+ {{\bf q^4t}+{\bf q^4t^2}(1-{\bf q^4t^2})} 
+{\bf q^6t^3}(1+{\bf t})}{1-{\bf q^4t^2}}
\cdot\frac{1-({\bf q^8t^6})^{^{\frac{n {+}1}{4}}}}{ 1-{\bf q^8t^6} }
\right)
\ee
\be
{\cal K}_{ur}^{[4,n]} = \FF^{^{\frac{3(n+1)}{2}}}\cdot {\bf t}^{-2}
\cdot\left(-{\bf qt}+{\bf q}^{-1}
+\underline{(1+{\bf t})\cdot\frac{1+{\bf q^4t}}{\bf q^3}
\cdot\frac{1-({\bf q^{12}t^8})^{^{\frac{n+1}{4}}}}{(1-{\bf q^4t^2})(1-{\bf q^6t^4})}}\ -
\right. \nn \\
\left.
-  {\left(1+{\bf q^4t}\right)}\cdot\frac{ 1+{\bf q^2}(1-{\bf q^4t^2})+{\bf q^4t^2}+{\bf q^6t^4}+{\bf q^8t^3}+{\bf q^{10}t^5}
-{\bf q^{12}t^6}}
{{\bf q^7t}(1-{\bf q^4t^2})}
\cdot\frac{1-({\bf q^8t^6})^{^{\frac{n+1}{4}}}}{ 1-{\bf q^8t^6} }
\right)
\ee
Despite this is not evident from the formulas,
for $n>0$ (but not for $n< 0$)
these polynomials are pure positive or pure negative,
depending on parity of $\frac{n+1}{2}$.
As already mentioned, this time there is no relation like (\ref{invn3}):
\be
{\cal K}_{_{\Box}}^{[4,-n]}({\bf q},{\bf t})  \ \ {/\!\!\!\!\!\!\sim}\
{\cal K}_{_{_\Box}}^{[4,n]}\left(\frac{1}{\bf q},\frac{1}{\bf t}\right)
\ee
neither in {\it unreduced} nor in {\it reduced} case,
moreover, the l.h.s. is not even a positive Laurent polynomial.
The valid topological identity is
\be
{\cal K}_{_{\Box}}^{[4,3]} =  {\cal K}_{_{\Box}}^{[3,4]}
\ee

Also, it is unclear how
the rule (\ref{redqn}) can be applied to derive {\it unreduced} polynomials
from the {\it reduced} ones.
This  drawback is cured by an
improved (better structured) version of above formulas --
which one day should be directly deduced from an adequately structured
expression for the superpolynomials.

For $n = 1\, {\rm mod}\,4$:
{\footnotesize
\be
{\cal K}_{r}^{[4,n]} = \FF^{^{\frac{3(n-1)}{2}}}\cdot\left(1+
(1+{\bf q^2t})\cdot{\bf q^4t^2}(1+{\bf q^8t^6})\cdot
\frac{1-({\bf q^8t^6})^{^{\frac{n-1}{4}}}}{ 1-{\bf q^8t^6} }
+ (1+{\bf q^4t})\cdot{\bf q^6t^4}(1+{\bf q^2t^2})\cdot
\frac{1-({\bf q^8t^6})^{^{\frac{n-1}{4}}}}{ 1-{\bf q^8t^6} } +
\right. \nn \\ \left.
+ (1+{\bf q^2t})\cdot\frac{{\bf q^{18}t^{11}}(1+{\bf t})}{1-{\bf q^6t^4}}\cdot
\frac{1-({\bf q^8t^6})^{^{\frac{n-1}{4}}}}{ 1-{\bf q^8t^6} }
- (1+{\bf q^2t})\cdot\frac{{\bf q^{10}t^{5}}(1+{\bf t})}{1-{\bf q^6t^4}}\cdot
\frac{1-({\bf q^4t^2})^{^{\frac{n-1}{4}}}}{ 1-{\bf q^4t^2} }\cdot
 ({\bf q^8t^6})^{^{\frac{n-1}{4}}}
 \right)
\nn
\ee
}

\noindent
Together the two items in the second line  form a positive polynomial,
but this positivity is still not explicit.
But now the {\it unreduced} polynomial is obtained by direct application of (\ref{reduct2}):
{\footnotesize
\be
{\cal K}_{ur}^{[4,n]} = \FF^{^{\frac{3(n-1)}{2}}}\cdot\left([2]+
(1+{\bf q^4t})\cdot{\bf q^3t^2}(1+{\bf q^8t^6}\cdot
\frac{1-({\bf q^8t^6})^{^{\frac{n-1}{4}}}}{ 1-{\bf q^8t^6} }
+ [2]\cdot (1+{\bf q^4t})\cdot{\bf q^6t^4}(1+{\bf q^2t^2})\cdot
\frac{1-({\bf q^8t^6})^{^{\frac{n-1}{4}}}}{ 1-{\bf q^8t^6} } +
\right. \nn \\ \left.
+ (1+{\bf q^4t})\cdot\frac{{\bf q^{17}t^{11}}(1+{\bf t})}{1-{\bf q^6t^4}}\cdot
\frac{1-({\bf q^8t^6})^{^{\frac{n-1}{4}}}}{ 1-{\bf q^8t^6} }
- (1+{\bf q^4t})\cdot\frac{{\bf q^{9}t^{5}}(1+{\bf t})}{1-{\bf q^6t^4}}\cdot
\frac{1-({\bf q^4t^2})^{^{\frac{n-1}{4}}}}{ 1-{\bf q^4t^2} }\cdot
 ({\bf q^8t^6})^{^{\frac{n-1}{4}}}
 \right)
\nn
\ee
}

\noindent
The analogues of these formulas for $n=-1\,{\rm mod}\,4$ are a little more complicated:
{\footnotesize
\be
{\cal K}_{r}^{[4,n]} = \FF^{^{\frac{3(n-1)}{2}}} \cdot\left(1+
\Big((1+{\bf q^2t})\cdot{\bf q^4t^2} + (1+{\bf q^4t})\cdot{\bf q^6t^4}\Big)\cdot
\frac{1-({\bf q^8t^6})^{^{\frac{n+1}{4}}}}{ 1-{\bf q^8t^6} }
+ \right. \nn \\ \left.
+ \Big((1+{\bf q^2t})\cdot{\bf q^{12}t^8} + (1+{\bf q^4t})\cdot{\bf q^8t^6}\Big)\cdot
\frac{1-({\bf q^8t^6})^{^{\frac{n-3}{4}}}}{ 1-{\bf q^8t^6} } +
\right. \nn \\ \left.
+ (1+{\bf q^2t})\cdot\frac{{\bf q^{18}t^{11}}(1+{\bf t})}{1-{\bf q^6t^4}}\cdot
\frac{1-({\bf q^8t^6})^{^{\frac{n-3}{4}}}}{ 1-{\bf q^8t^6} }
- (1+{\bf q^2t})\cdot\frac{{\bf q^{8}t^{3}}(1+{\bf t})}{1-{\bf q^6t^4}}\cdot
\frac{1-({\bf q^4t^2})^{^{\frac{n-3}{4}}}}{ 1-{\bf q^4t^2} }\cdot
 ({\bf q^8t^6})^{^{\frac{n+1}{4}}}
 \right)
\nn
\ee
}
and
{\footnotesize
\be
\!\!\!\!\!\!\!\!
{\cal K}_{ur}^{[4,n]} = \FF^{^{\frac{3(n-1)}{2}}}  \cdot\left([2]+
(1+{\bf q^4t})\cdot\Big({\bf q^3t^2} + [2] \cdot{\bf q^6t^4}\Big)\cdot
\frac{1-({\bf q^8t^6})^{^{\frac{n+1}{4}}}}{ 1-{\bf q^8t^6} }
+ \ (1+{\bf q^4t})\cdot\Big(  {\bf q^{11}t^8} + [2]\cdot{\bf q^8t^6}\Big)\cdot
\frac{1-({\bf q^8t^6})^{^{\frac{n-3}{4}}}}{ 1-{\bf q^8t^6} } \ +
\right.
\nn
\ee
\vspace{-0.4cm}
\be
\left.
+ \ (1+{\bf q^4t})\cdot\frac{{\bf q^{17}t^{11}}(1+{\bf t})}{1-{\bf q^6t^4}}\cdot
\frac{1-({\bf q^8t^6})^{^{\frac{n-3}{4}}}}{ 1-{\bf q^8t^6} }
- (1+{\bf q^4t})\cdot\frac{{\bf q^{7}t^{3}}(1+{\bf t})}{1-{\bf q^6t^4}}\cdot
\frac{1-({\bf q^4t^2})^{^{\frac{n-3}{4}}}}{ 1-{\bf q^4t^2} }\cdot
 ({\bf q^8t^6})^{^{\frac{n+1}{4}}}
 \right)
\nn
\ee
}

\section{The case of $m=5$}

This time there are four different kinds of formulas,
for four different residues $n = 1,2,3,4\, {\rm mod}\, 5$.
The corresponding formulas for  Khovanov polynomials are:

\bigskip

{\bf Residue 1}
{\footnotesize
\be
{\cal K}^{[5,n]}_r =\FF^{^{2(n-1)}}\cdot\left(1
-\frac{{\bf q^{10}t^5}(1+{\bf q^8t^6})\Big((1+{\bf q^2t})(1+{\bf t}+{\bf q^2t^3}+{\bf q^6t^4})
+(1+{\bf q^4t}){\bf q^2t^2}(1+{\bf q^2t^3})
\Big)}{1-{\bf q^6t^4}}\cdot
\frac{1-({\bf q^{16}t^{12}})^{^{\frac{n-1}{4}}}}{ 1-{\bf q^{16}t^{12}} }
+ \right. \nn \\ \left.
\!\!\!\!\!\!\!\!\!\!\!\!\!\!\!\!\!\!\!\!\!
+\frac{(1+{\bf q^2t})\cdot{\bf q^4t^2}(1+{\bf q^6t^3} +2{\bf q^8t^6}+{\bf q^{10}t^8}
+{\bf q^{12}t^7} +2{\bf q^{14}t^9} )
+(1+{\bf q^4t})\cdot{\bf q^6t^4}(1+{\bf q^2t^2} + {\bf q^4t^4} + {\bf q^6t^3}-{\bf q^6t^4}+{\bf q^{14}t^9})
}{1-{\bf q^6t^4}}\cdot
\frac{1-({\bf q^{10}t^8})^{^{\frac{n-1}{4}}}}{ 1-{\bf q^{10}t^8} }
 \right)
\nn \\ \nn \\ \nn \\
{\cal K}^{[5,n]}_{ur} =\FF^{^{2(n-1)}}\cdot\left([2]
-\frac{{\bf q^{9}t^5}(1+{\bf q^8t^6})(1+{\bf q^4t})
\Big( \overbrace{(1+{\bf t}+{\bf q^2t^3}+{\bf q^6t^4})
+ [2]\cdot {\bf q^3t^2}(1+{\bf q^2t^3})}^{(1+{\bf t})\cdot
\Big(1+{\bf q^2t^2} + {\bf q^4t^2}(1-{\bf t}+{\bf t^2})+ {\bf q^6t^6}\Big)}
\Big)}{1-{\bf q^6t^4}}\cdot
\frac{1-({\bf q^{16}t^{12}})^{^{\frac{n-1}{4}}}}{ 1-{\bf q^{16}t^{12}} }
+ \right. \nn \\ \left.
\!\!\!\!\!\!\!\!\!\!\!\!\!
+(1+{\bf q^4t})\cdot\frac{{\bf q^3t^2}(1+{\bf q^6t^3} +2{\bf q^8t^6}+{\bf q^{10}t^8}
+{\bf q^{12}t^7} +2{\bf q^{14}t^9} )
+[2]\cdot {\bf q^6t^4}(1+{\bf q^2t^2} + {\bf q^4t^4} + {\bf q^6t^3}-{\bf q^6t^4}+{\bf q^{14}t^9})
}{1-{\bf q^6t^4}}\cdot
\frac{1-({\bf q^{10}t^8})^{^{\frac{n-1}{4}}}}{ 1-{\bf q^{10}t^8} }
 \right)
\nn
\ee
}

\bigskip

{\bf Residue 2}
{\footnotesize
\be
{\cal K}^{[5,n]}_r =\FF^{^{2(n-1)}}\!\!\!\cdot\!\!\left(1
+ {\bf q^4t^2}(1+{\bf q^2t})(1+{\bf q^4t^2}) -
\phantom{\frac{1-({\bf q^{10]t^8})^{^{\frac{n-2}{4}}}}}{ 1-{\bf q^{10}t^8} }}
\right. \nn \\ \left.
-\frac{{\bf q^{12}t^7}(1+{\bf q^8t^6})
\Big((1+{\bf q^2t})\Big[(1+{\bf t})({\bf q^{10}t^6} + {\bf q^6t^4}-{\bf q}^2)
+{\bf t}(1+{\bf q^4t})\Big] + (1+{\bf q^4t})(1+{\bf q^2t^3}) \Big)}{1-{\bf q^6t^4}}\cdot
\frac{1-({\bf q^{16}t^{12}})^{^{\frac{n-2}{4}}}}{ 1-{\bf q^{16}t^{12}} }
+ \right. \nn \\ \left.
\!\!\!\!\!\!\!\!\!\!\!\!\!\!\!\!\!\!\!\!\!
+\left[\frac{(1+{\bf q^2t})\cdot{\bf q^8t^4}(-1+2{\bf q^4t^4}-{\bf q^6t^3} +2{\bf q^6t^6}
+{\bf q^8t^5}+2{\bf q^{10}t^7}+{\bf q^{10}t^8}
+{\bf q^{12}t^9} + 
 {\bf q^{16}t^{11}} )}{1-{\bf q^6t^4}}
+ \ \ \ \ \ \ \ \ \ \ \ \ \ \ \ \ \ \ \ \ \ \ \ \ \ \ \right.\right.\nn \\ \left.\left.
+ \frac{(1+{\bf q^4t})\cdot{\bf q^6t^4}(1+{\bf q^2t^2} + {\bf q^4t^4}
+ {\bf q^6t^3}-{\bf q^6t^4}+{\bf q^{14}t^9})}{1-{\bf q^6t^4}}
\right]\cdot
\frac{1-({\bf q^{10}t^8})^{^{\frac{n-2}{4}}}}{ 1-{\bf q^{10}t^8} }
 \right)
\nn \\ \nn \\ \nn \\
{\cal K}^{[5,n]}_{ur} =\FF^{^{2(n-1)}}\cdot\left([2]
+ {\bf q^3t^2}(1+{\bf q^4t})(1+{\bf q^4t^2}) -
\phantom{\frac{1-({\bf q^{10]t^8})^{^{\frac{n-2}{4}}}}}{ 1-{\bf q^{10}t^8} }}
 \right. \nn \\ \left.
-\frac{{\bf q^{11}t^7}(1+{\bf q^8t^6})(1+{\bf q^4t})
\Big(\Big[(1+{\bf t})({\bf q^{10}t^6} + {\bf q^6t^4}-{\bf q}^2)
+{\bf t}(1+{\bf q^4t})\Big] +[2]\cdot {\bf q}(1+{\bf q^2t^3}) \Big)}{1-{\bf q^6t^4}}\cdot
\frac{1-({\bf q^{16}t^{12}})^{^{\frac{n-2}{4}}}}{ 1-{\bf q^{16}t^{12}} }
+  \right. \nn \\ \left.
\!\!\!\!\!\!\!\!\!\!\!\!\!\!\!\!\!\!\!\!\!
+(1+{\bf q^4t})\cdot\left[\frac{{\bf q^7t^4}(-1+2{\bf q^4t^4}-{\bf q^6t^3} +2{\bf q^6t^6}
+{\bf q^8t^5}+2{\bf q^{10}t^7}+{\bf q^{10}t^8}
+{\bf q^{12}t^9} + 
 {\bf q^{16}t^{11}} )}{1-{\bf q^6t^4}}
+ \ \ \ \ \ \ \ \ \ \ \ \ \ \ \ \ \ \ \ \ \ \ \ \ \ \  \right.\right.\nn \\ \left.\left.
+ \frac{[2]\cdot{\bf q^6t^4}(1+{\bf q^2t^2} + {\bf q^4t^4}
+ {\bf q^6t^3}-{\bf q^6t^4}+{\bf q^{14}t^9})}{1-{\bf q^6t^4}}
\right]\cdot
\frac{1-({\bf q^{10}t^8})^{^{\frac{n-2}{4}}}}{ 1-{\bf q^{10}t^8} }
 \right)
\nn
\ee
}

\bigskip

{\bf Residue 3}
{\footnotesize
\be
{\cal K}^{[5,n]}_r =\FF^{^{2(n-1)}}\!\!\!\cdot\!\!\left(1
+ (1+{\bf q^2t})\cdot{\bf q^4t^2}(1+{\bf q^6t^4}) +(1+{\bf q^4t})\cdot{\bf q^6t^4}-
\phantom{\frac{1-({\bf q^{10]t^8})^{^{\frac{n-3}{4}}}}}{ 1-{\bf q^{10}t^8} }}
\right. \nn \\ \left. \!\!\!\!\!\!\!\!\!\!\!\!\!\!\!\!\!
-\frac{\Big((1+{\bf q^2t})\cdot (1-{\bf q^4t^2}+{\bf q^6t^3}
+ {\bf q^6t^4}+{\bf q^{12}t^7}+{\bf q^{12}t^8})
  + (1+{\bf q^4t})\cdot (-1+{\bf q^2t^2}+{\bf q^8t^5}+{\bf q^8t^6}) \Big)
  \cdot {\bf q^{12}t^8}(1+{\bf q^8t^6})}{1-{\bf q^6t^4}}\cdot
\frac{1-({\bf q^{16}t^{12}})^{^{\frac{n-3}{4}}}}{ 1-{\bf q^{16}t^{12}} }
+ \right. \nn \\ \left.
+\left[\frac{(1+{\bf q^2t})\cdot{\bf q^{10}t^6}(-1+2{\bf q^2t^2}+2{\bf q^4t^4}
 +2{\bf q^8t^5}+{\bf q^{10}t^7}+{\bf q^{10}t^8}+{\bf q^{16}t^{11}} )}{1-{\bf q^6t^4}}
+ \ \ \ \ \ \ \ \ \ \ \ \ \ \ \ \ \ \ \ \ \ \ \ \ \ \
\ \ \ \ \ \ \ \ \ \ \ \ \  \right.\right.\nn \\ \left.\left.
+ \frac{(1+{\bf q^4t})\cdot{\bf q^8t^6}(1+{\bf q^2t^2})(1+{\bf q^2t})(1-{\bf q^2t} + {\bf q^6t^4}
 -{\bf q^8t^5}+{\bf q^{10}t^6})}{1-{\bf q^6t^4}}
\right]\cdot
\frac{1-({\bf q^{10}t^8})^{^{\frac{n-3}{4}}}}{ 1-{\bf q^{10}t^8} }
 \right)
\nn \\ \nn \\ \nn \\
{\cal K}^{[5,n]}_{ur} =\FF^{^{2(n-1)}}\cdot\left([2]
+ (1+{\bf q^4t})\cdot\Big({\bf q^3t^2}(1+{\bf q^6t^4}) +[2]\cdot{\bf q^6t^4}\Big)-
\phantom{\frac{1-({\bf q^{10]t^8})^{^{\frac{n-3}{4}}}}}{ 1-{\bf q^{10}t^8} }}
\right. \nn \\ \left. \!\!\!\!\!\!\!\!\!\!\!\!\!\!\!\!\!
-(1+{\bf q^4t})\cdot\frac{\Big( (1-{\bf q^4t^2}+{\bf q^6t^3}
+ {\bf q^6t^4}+{\bf q^{12}t^7}+{\bf q^{12}t^8})
  + [2]\,{\bf q}\cdot (-1+{\bf q^2t^2}+{\bf q^8t^5}+{\bf q^8t^6}) \Big)
  \cdot {\bf q^{11}t^8}(1+{\bf q^8t^6})}{1-{\bf q^6t^4}}\cdot
\frac{1-({\bf q^{16}t^{12}})^{^{\frac{n-3}{4}}}}{ 1-{\bf q^{16}t^{12}} }
+ \right. \nn \\ \left.
+(1+{\bf q^4t})\cdot\left[\frac{{\bf q^{9}t^6}(-1+2{\bf q^2t^2}+2{\bf q^4t^4}
 +2{\bf q^8t^5}+{\bf q^{10}t^7}+{\bf q^{10}t^8}+{\bf q^{16}t^{11}} )}{1-{\bf q^6t^4}}
+ \ \ \ \ \ \ \ \ \ \ \ \ \ \ \ \ \ \ \ \ \ \ \ \ \ \
\ \ \ \ \ \ \ \ \ \ \ \ \  \right.\right.\nn \\ \left.\left.
+ [2]\cdot\frac{{\bf q^8t^6}(1+{\bf q^2t^2})(1+{\bf q^2t})(1-{\bf q^2t} + {\bf q^6t^4}
 -{\bf q^8t^5}+{\bf q^{10}t^6})}{1-{\bf q^6t^4}}
\right]\cdot
\frac{1-({\bf q^{10}t^8})^{^{\frac{n-3}{4}}}}{ 1-{\bf q^{10}t^8} }
 \right)
\nn
\ee
}

\bigskip

{\bf Residue 4}
{\footnotesize
\be
{\cal K}^{[5,n]}_r =\FF^{^{2(n-1)}}\!\!\!\cdot\!\!\left(1
+ (1+{\bf q^2t})\cdot{\bf q^4t^2}(1+{\bf q^8t^6})
+(1+{\bf q^4t})\cdot{\bf q^6t^4}(1+{\bf q^2t^2})-
\phantom{\frac{1-({\bf q^{10]t^8})^{^{\frac{n-4}{4}}}}}{ 1-{\bf q^{10}t^8} }}
\right. \nn \\ \left. \!\!\!\!\!\!\!\!\!\!\!\!\!\!\!\!\!
-\frac{{\bf q^{20}t^{13}}(1+{\bf q^8t^6})\cdot
\Big(\overbrace{(1+{\bf q^2t})\cdot  {\bf t}\cdot(1+{\bf q^4t }+{\bf q^6t^3}
+ {\bf q^6t^4} )
  + (1+{\bf q^4t})\cdot (1+{\bf q^2t^3})}^{(1+{\bf t})\cdot
 \Big( (1+{\bf q^2t})\cdot{\bf q^6t^4}+(1+{\bf q^4t})\cdot(1+{\bf q^2t^2})  \Big) } \Big) }
{1-{\bf q^6t^4}}\cdot
\frac{1-({\bf q^{16}t^{12}})^{^{\frac{n-4}{4}}}}{ 1-{\bf q^{16}t^{12}} }
+ \right. \nn \\ \left.
+\left[\frac{(1+{\bf q^2t})\cdot{\bf q^{14}t^{10}}(1+{\bf q^8t^6})
\Big(2+{\bf q^4t}+{\bf q^6t^3}\Big) }
{1-{\bf q^6t^4}}
+ \ \ \ \ \ \ \ \ \ \ \ \ \ \ \ \ \ \ \ \ \ \ \ \ \ \
\ \ \ \ \ \ \ \ \ \ \ \ \  \right.\right.\nn \\ \left.\left.
+ \frac{(1+{\bf q^4t})\cdot{\bf q^{10}t^8} (1+{\bf q^2t})\Big(
(1-{\bf q^2t})(1+{\bf q^4t^2} + {\bf q^6t^4}+{\bf q^8t^6})
+{\bf q^{8}t^4}(1+{\bf q^2t^2})\Big) }{1-{\bf q^6t^4}}
\right]\cdot
\frac{1-({\bf q^{10}t^8})^{^{\frac{n-4}{4}}}}{ 1-{\bf q^{10}t^8} }
 \right)
\nn \\ \nn \\ \nn \\
{\cal K}^{[5,n]}_{ur} =\FF^{^{2(n-1)}}\cdot\left([2]
+ (1+{\bf q^4t})\cdot\Big({\bf q^3t^2}(1+{\bf q^8t^6}) +[2]\cdot{\bf q^6t^4}(1+{\bf q^2t^2})\Big)-
\phantom{\frac{1-({\bf q^{10]t^8})^{^{\frac{n-4}{4}}}}}{ 1-{\bf q^{10}t^8} }}
\right. \nn \\ \left. \!\!\!\!\!\!\!\!\!\!\!\!\!\!\!\!\!
-(1+{\bf q^4t})\cdot\frac{{\bf q^{19}t^{13}}(1+{\bf q^8t^6})\cdot
\Big(  {\bf t}\cdot(1+{\bf q^4t }+{\bf q^6t^3}
+ {\bf q^6t^4} )
  +  [2]\,{\bf q}\cdot (1+{\bf q^2t^3}) \Big)
  \cdot }{1-{\bf q^6t^4}}
\frac{1-({\bf q^{16}t^{12}})^{^{\frac{n-4}{4}}}}{ 1-{\bf q^{16}t^{12}} }
+ \right. \nn \\ \left.
+(1+{\bf q^4t})\cdot
\left[\frac{ {\bf q^{13}t^{10}}(1+{\bf q^8t^6})\Big(2+{\bf q^4t}+{\bf q^6t^3}\Big) }
{1-{\bf q^6t^4}}
+ \ \ \ \ \ \ \ \ \ \ \ \ \ \ \ \ \ \ \ \ \ \ \ \ \ \
\ \ \ \ \ \ \ \ \ \ \ \ \  \right.\right.\nn \\ \left.\left.
+ \frac{[2] \cdot {\bf q^{10}t^8}(1+{\bf q^2t})\Big(
(1-{\bf q^2t})(1+{\bf q^4t^2} + {\bf q^6t^4}+{\bf q^8t^6})
+{\bf q^{8}t^4}(1+{\bf q^2t^2})\Big) }{1-{\bf q^6t^4}}
\right]\cdot{\bf q^{10}t^8}\cdot
\frac{1-({\bf q^{10}t^8})^{^{\frac{n-4}{4}}}}{ 1-{\bf q^{10}t^8} }
 \right)
\nn
\ee
}

\bigskip

\noindent
Some of these formulas could be considerably simplified,
as shown in a couple of overbraced
examples,
but we put them in the form
with explicit separation of $(1+{\bf q^2t})$ and $(1+{\bf q^4t})$ structures,
when relation between {\it reduced} and {\it unreduced} polynomials is
provided by the rule (\ref{reduct2}).
It is this structure that should be revealed in order to understand
the $N$-reduction of the superpolynomials \cite{AnoMredP}.

One can easily check the
$[m,n]\longleftrightarrow [n,m]$ equivalence:
\be
{\cal K}^{[5,2]} ={\cal K}^{[2,5]}, \ \ \ \ \ \
 {\cal K}^{[5,3]} ={\cal K}^{[3,5]}, \ \ \ \ \ \
{\cal K}^{[5,4]} ={\cal K}^{[4,5]}, \ \ \ \ \ \
\ee
while $n\longrightarrow -n$ relation fails:
\be
{\cal K}^{[5,-n]}({\bf q},{\bf t}) \ {/\!\!\!\!\!\!\sim}\
{\cal K}^{[5,n]}({\bf q}^{-1},{\bf t}^{-1})
\ee
moreover, the l.h.s. is not a positive polynomial.

\section{Projective limit}

If one throws away all the $n$-power factors from the formulas for ${\cal K}$,
i.e., keep only the contribution of the Young diagram $Q_0=[m]$,
one gets the {\it projective limit} of Khovanov polynomials, studied in \cite{Gprojlim},
which are positive {\it series}.
Explicit expressions are easily reproduced from our general formulas:

{
\be
{_{r}{K}}^{[2,\infty]}_{_\Box} = 1+\frac{\bf q^4 t^2 }{1-{\bf q^2t}}
= \frac{1+{\bf q^6t^3}}{ 1-{\bf q^4t^2}}
= 1+(1+{\bf q^2t})\cdot\frac{\bf q^4 t^2 }{1-{\bf q^4t^2}}
\nn\\
 {_{ur}{K}}^{[2,\infty]}_{_\Box}
= {\bf q} + \frac{1}{\bf q}\cdot\frac{1+{\bf q^8t^3}}{1-{\bf q^4t^2}}
= [2] +(1+{\bf q^4 t})\cdot\frac{{\bf q^3 t^2} }{1-{\bf q^4t^2}}
\nn \\ \nn \\ \nn \\
{_{r}{K}}^{[3,\infty]}_{_\Box}
= \frac{(1+{\bf q^4t^2})(1+{\bf q^6t^3})}{ 1-{\bf q^6t^4}}
= 1+ (1+{\bf q^2t})\cdot \frac{\bf q^4t^2}{1-{\bf q^6t^4}}
+ (1+{\bf q^4t})\cdot\frac{{\bf q^{6}t^4}}{1-{\bf q^6t^4}}
\nn\\
 {_{ur}{K}}^{[3,\infty]}_{_\Box}
= \frac{1}{\bf q}\cdot\frac{1+{\bf q^2}+{\bf q^4t^2}+{\bf q^8t^3}+{\bf q^{10}t^5}+{\bf q^{12}t^5}}{1-{\bf q^6t^4}}
= [2] + (1+{\bf q^4t})\cdot \frac{\bf q^3t^2}{1-{\bf q^6t^4}}
+[2]\cdot(1+{\bf q^4t})\cdot\frac{{\bf q^{6}t^4}}{1-{\bf q^6t^4}}
\nn \\ \nn \\ \nn \\
{_{r}{K}}^{[4,\infty]}_{_\Box}
= 1 + (1+{\bf q^2t})\cdot\frac{{\bf q^4t^2}(1+{\bf q^8t^6})}{ 1-{\bf q^8t^6} }
+ (1+{\bf q^4t})\cdot\frac{{\bf q^6t^4}(1+{\bf q^2t^2})}{1-{\bf q^8t^6}}
+ (1+{\bf q^2t})\cdot\frac{{\bf q^{18}t^{11}}(1+{\bf t})}{(1-{\bf q^6t^4})( 1-{\bf q^8t^6}) }
\nn\\
 {_{ur}{K}}^{[4,\infty]}_{_\Box}
= [2] + (1+{\bf q^4t})\cdot\frac{{\bf q^3t^2}(1+{\bf q^8t^6})}{ 1-{\bf q^8t^6} }
+ [2]\cdot(1+{\bf q^4t})\cdot\frac{{\bf q^6t^4}(1+{\bf q^2t^2})}{1-{\bf q^8t^6}}
+ (1+{\bf q^4t})\cdot\frac{{\bf q^{17}t^{11}}(1+{\bf t})}{(1-{\bf q^6t^4})( 1-{\bf q^8t^6}) }
\nn \\ \nn \\ \nn \\
{_{r}{K}}^{[5,\infty]}_{_\Box}
= 1 + (1+{\bf q^2t})\cdot\frac{{\bf q^4t^2}}{ 1-{\bf q^{10}t^8} }
\cdot \left(1+{\bf q^8t^6}+\frac{{\bf q^{10}t^8}(1+{\bf q^4t})}{1-{\bf q^8t^6}}\right)
+\nn \\
+ (1+{\bf q^4t})\cdot\frac{{\bf q^6t^4}}{1-{\bf q^{10}t^8}}
\cdot \left(1+{\bf q^2t^2}+{\bf q^4t^4}
+\frac{{\bf q^{10}t^8}(1+{\bf q^6t^3})}{1-{\bf q^8t^6}} \right)
+ (1+{\bf q^2t})\cdot\frac{{\bf q^{20}t^{13}}(1+{\bf t})}{(1-{\bf q^8t^6})( 1-{\bf q^{10}t^8}) }
\nn\\
 {_{ur}{K}}^{[5,\infty]}_{_\Box}
 = [2] + (1+{\bf q^4t})\cdot\frac{{\bf q^3t^2}}{ 1-{\bf q^{10}t^8} }
\cdot \left(1+{\bf q^8t^6}+\frac{{\bf q^{10}t^8}(1+{\bf q^4t})}{1-{\bf q^8t^6}}\right)
+\nn \\
+[2]\cdot (1+{\bf q^4t})\cdot\frac{{\bf q^6t^4}}{1-{\bf q^{10}t^8}}
\cdot \left(1+{\bf q^2t^2}+{\bf q^4t^4}
+\frac{{\bf q^{10}t^8}(1+{\bf q^6t^3})}{1-{\bf q^8t^6}} \right)
+ (1+{\bf q^4t})\cdot\frac{{\bf q^{19}t^{13}}(1+{\bf t})}{(1-{\bf q^8t^6})( 1-{\bf q^{10}t^8}) }
\nn
\ee
\be
\ldots
\nn
\ee
}
The difference between  branches disappears in this limit.
{\it Unreduced} polynomials are obtained from {\it reduced} ones by
multiplication by $[2]$ and application of the rule (\ref{reduct2}):
$[2]\cdot (1+{\bf q^2t}) \cong \frac{1}{\bf q}\cdot (1+{\bf q^4t})$.
Despite $K_{_\Box}^{[m,\infty]}$ are nothing but Khovanov's substitutes
of the nicely factorized MacDonald dimensions $M_{[m]}^*$ at $N=2$,
they fail to {\it fully}  factorize, starting already from $m=4$:
{
\be
{_{r}{K}}^{[2,\infty]}_{_\Box} = \frac{1+{\bf q^6t^3}}{ 1-{\bf q^4t^2}}\nn \\
{_{r}{K}}^{[3,\infty]}_{_\Box}
= \frac{(1+{\bf q^6t^3})}{ 1-{\bf q^6t^4}}\cdot\Big(1+{\bf q^4t^2}\Big) \nn \\
{_{r}{K}}^{[4,\infty]}_{_\Box} =
\frac{1+{\bf q^6t^3}}{(1-{\bf q^6t^4})(  1-{\bf q^8t^6})  }\cdot
\Big(1+{\bf q^4t^2}(1-{\bf q^6t^4})+ {\bf q^{14}t^9}\Big)\nn \\
{_{r}{K}}^{[5,\infty]}_{_\Box} = \frac{1+{\bf q^6t^3}}{  (1-{\bf q^8t^6})(1-{\bf q^{10}t^8})}
\cdot\Big(1+{\bf q^4t^2} +{\bf q^6t^4}+  {\bf q^{14}t^9}+{\bf q^{16}t^{11}}
+{\bf q^{20}t^{13}}\Big)\nn \\
{_{r}{K}}^{[6,\infty]}_{_\Box} =
\frac{1+{\bf q^6t^3}}{  (1-{\bf q^8t^6})(1-{\bf q^{10}t^8})(1-{\bf q^{12}t^{10}})}\cdot
\ \ \ \ \ \ \ \ \ \ \ \ \ \ \ \ \ \ \ \ \ \ \ \ \ \ \ \ \ \ \ \ \ \ \ \ \ \
\ \ \ \ \ \ \ \ \ \ \ \ \ \ \ \ \ \ \ \ \ \ \ \ \
\\
\cdot\Big(1+({\bf q^4t^2}+  {\bf q^{14}t^9})(1 -{\bf q^{16}t^{12}} )
+({\bf q^6t^4}+{\bf q^{20}t^{13}})(1-{\bf q^8t^6})
+{\bf q^{16}t^{11}}+{\bf q^{18}t^{13}}+{\bf q^{34}t^{24}}\Big) \nn \\
\ldots \nn
\ee
}
Formulas are written in the form, where positivity of the series is explicitly seen.

\section{ Recursion for KR polynomials at $N=3$ and $m=4$
\label{N3}}

KR polynomial with $N=3$ coincides with the superpolynomial at $m=2,3$,
and deviates from it for $m\geq 4$.
For $m=4$ four out of five terms in the evolution formula will contribute,
and the four $n$-independent coefficients can be found from the known
expressions for particular polynomials computed with~\cite{sl3prog}.
For {\it reduced} case  we get for  $n=1 \,{\rm mod}\, 4$:

{\footnotesize
\be
_r {\cal K}^{[4,n]}_{_\Box}(N=3) = \FFF^{^\frac{3(n-1)}{2}}\cdot\left(1
+  \frac{{\bf q^4t^2}(1+{\bf q^8t^3})}{(1-{\bf q^4t^2})(1-{\bf q^8t^4})}\cdot
\Big( 1+{\bf q^2t^2} + {\bf q^4t} + {\bf q^4t^4} + {\bf q^6t^3} + {\bf q^6t^4}
- \ \ \ \ \ \ \ \ \ \ \ \ \ \ \right.\nn \\ \left.
- {\bf q^8t^4} - {\bf q^{10}t^4} + {\bf q^{10}t^5} - {\bf q^{12}t^6}
-{\bf q^{14}t^5}-{\bf q^{14}t^8} - {\bf q^{16} t^7} + {\bf q^{16}t^9}
-{\bf q^{18}t^9} - {\bf q^{20}t^9}
\Big)
\cdot\frac{1-({\bf q^8t^6})^{^\frac{n-1}{4}}}{1-{\bf q^8t^6}}
+ \right.\nn \\ \left.
+ \frac{(1-{\bf t}^2)\cdot {\bf q^{10}t^4}(1+{\bf q^4t}) (1+{\bf q^8t^3})}
{(1-{\bf q^4t^2})^2}
\cdot\frac{1-({\bf q^{12}t^8})^{^\frac{n-1}{4}}}{1-{\bf q^{6}t^4}}
- \underbrace{{\bf q^{10}t^4}(1+{\bf q^4t})(1+{\bf q^2t} + {\bf q^2t^2} + {\bf q^8t^3})}_
{\!\!\!\!\!\!\!\!\!\!\!
{\bf q^{10}t^4}(1+{\bf q^4t})\Big((1+{\bf q^2t})+{\bf q^2t^2}(1+{\bf q^6t})\Big)
\!\!\!\!\!\!\!\!\!\!\!}
\cdot\frac{1-({\bf q^{16}t^{10}})^{^\frac{n-1}{4}}}{(1-{\bf q^4t^2})(1-{\bf q^{8}t^{4})}
}\right)
\nn
\ee
}
and
{\footnotesize
\be
_{ur} {\cal K}^{[4,n]}_{_\Box}(N=3) = \FFF^{^\frac{3(n-1)}{2}}\cdot\left([3]
+  \frac{{\bf q^2t^2}(1+{\bf q^6t})}{(1-{\bf q^4t^2})(1-{\bf q^8t^4})}\cdot
\Big( 1+{\bf q^2} +{\bf q^2t^2} + {\bf q^4t^2} + {\bf q^4t^4} + 2 {\bf q^6t^4}
+ {\bf q^8t^3}+{\bf q^8t^4} +{\bf q^{10}t^3}
-   \right.\nn \\ \left.
-2 {\bf q^{10}t^4}
+ 2{\bf q^{12}t^5}
- 2{\bf q^{12}t^6}-{\bf q^{14}t^6}
+   \right.\nn \\ \left.
 +2{\bf q^{14}t^7}-{\bf q^{14}t^8} + {\bf q^{16} t^7} -2 {\bf q^{16}t^8}-{\bf q^{18}t^7}
+{\bf q^{18}t^9} -{\bf q^{20}t^7}- {\bf q^{20}t^9} - {\bf q^{22}t^9}-{\bf q^{22}t^{11}}
- 2{\bf q^{24}t^{11}}-{\bf q^{26}t^{11}}
\Big)
\cdot\frac{1-({\bf q^8t^6})^{^\frac{n-1}{4}}}{1-{\bf q^8t^6}}
+ \right.\nn \\ \left.
+ (1+{\bf q^6t})\cdot
\frac{(1+{\bf t})\cdot {\bf q^{8}t^4}  }
{(1-{\bf q^4t^2})^2}
\cdot\Big(1-{\bf t} - {\bf q^2t} + {\bf q^8t^3} +{\bf q^{10}t^3}-{\bf q^{10}t^4}\Big)
\cdot\frac{(1-{\bf q^{12}t^8})^{^\frac{n-1}{4}}}{1-{\bf q^{6}t^4}}
- \right.\nn \\
- \underbrace{{\bf q^{8}t^4}(1+{\bf q^6t})
\Big(1+{\bf q^2t^2} + {\bf q^4t}+{\bf q^4t^2} + {\bf q^8t^3}+{\bf q^{10}t^3}\Big)}_
{{\bf q^{10}t^4}(1+{\bf q^6t})\Big(\frac{1}{\bf q^2}(1+{\bf q^4t})
+ \frac{1}{\bf q}[2]\cdot {\bf q^2t^2}(1+{\bf q^6t})\Big)}
\left.
\cdot\frac{1-({\bf q^{16}t^{10}})^{^\frac{n-1}{4}}}{(1-{\bf q^4t^2})(1-{\bf q^{8}t^{4})}
}\right)
\nn
\ee
}

\noindent
while for $n=-1 \,{\rm mod}\, 4$

{\footnotesize
\be
_r {\cal K}^{[4,n]}_{_\Box}(N=3) = \FFF^{^\frac{3(n-1)}{2}}\cdot\left( 1 +
{\bf q^4t^2}\cdot(1+{\bf q^4t})(1+{\bf q^2t^2} + {\bf q^4t^2} + {\bf q^8t^4} + {\bf q^{10}t^5})
%
+\frac{{\bf q^8t^6}(1+{\bf q^8t^3})}{(1-{\bf q^4t^2})(1-{\bf q^8t^4})}\cdot
\Big( 1+{\bf q^2 } - {\bf q^4 }
+ \right.\nn \\ \left.
+ {\bf q^4t^2} + {\bf q^6t} + {\bf q^6t^4}
+ {\bf q^8t^3} - {\bf q^{10}t^4} + {\bf q^{10}t^5} + {\bf q^{12}t^5}
-{\bf q^{14}t^5}-{\bf q^{14}t^6} - {\bf q^{16} t^5} - {\bf q^{16}t^8}
-{\bf q^{18}t^7} - {\bf q^{20}t^9}
\Big)
\cdot\frac{1-({\bf q^8t^6})^{^\frac{n-3}{4}}}{1-{\bf q^8t^6}}
+ \right.\nn \\
+ \frac{(1-{\bf t}^2)\cdot {\bf q^{16}t^8}(1+{\bf q^4t})(1+{\bf q^8t^3})}
{(1-{\bf q^4t^2})^2}
\cdot\frac{1-({\bf q^{12}t^8})^{^\frac{n-3}{4}}}{ {1-{\bf q^6t^4}}}
- \underbrace{{\bf q^{18}t^{10}}(1+{\bf q^4t})(1+{\bf q^6t} + {\bf q^6t^2} + {\bf q^8t^3})}_
{{\bf q^{18}t^{10}}\Big((1+{\bf q^4t})(1+{\bf q^6t})+{\bf q^6t^2}(1+{\bf q^2t})(1+{\bf q^4t})\Big)}
\left.
\cdot\frac{1-({\bf q^{16}t^{10}})^{^\frac{n-3}{4}}}{(1-{\bf q^4t^2})(1-{\bf q^{8}t^{4})}
}\right)
\nn
\ee
}
and
{\footnotesize
\be
_{ur} {\cal K}^{[4,n]}_{_\Box}(N=3) = \FFF^{^\frac{3(n-1)}{2}}\cdot\left([3]
+ (1+{\bf q^6t})\cdot {\bf q^2t^2}(1+{\bf q^2}+[2]^2{\bf q^4t^2} + {\bf q^8t^4}+{\bf q^{12}t^5})
+  \frac{{\bf q^8t^6}(1+{\bf q^6t})}{(1-{\bf q^4t^2})(1-{\bf q^8t^4})}\cdot
\Big( [3]
+   \right.\nn \\ \left.
+1+{\bf q^2t^2}  + {\bf q^4t^2}  +{\bf q^4t^4} -{\bf q^6t^2}+  {\bf q^6t^4}
+ 2{\bf q^8t^3}-{\bf q^8t^4} +{\bf q^{10}t^3}
-2 {\bf q^{10}t^4} + {\bf q^{10}t^5}
+ 2{\bf q^{12}t^5}
- 2{\bf q^{12}t^6}
+   \right.\nn \\ \left.
 +2{\bf q^{14}t^7}-{\bf q^{14}t^8} - {\bf q^{16} t^7} - {\bf q^{16}t^8}-2{\bf q^{18}t^7}
  -{\bf q^{20}t^7}- {\bf q^{20}t^9} - {\bf q^{22}t^9}-{\bf q^{22}t^{11}}
- {\bf q^{24}t^{11}}
\Big)
\cdot\frac{1-({\bf q^8t^6})^{^\frac{n-\blue{3}}{4}}}{1-{\bf q^8t^6}}
+ \right.\nn \\ \left.
+ (1+{\bf q^6t})\cdot
\frac{(1+{\bf t})\cdot {\bf q^{14}t^8} }
{(1-{\bf q^4t^2})^2}
\cdot\Big(1-{\bf t} - {\bf q^2t} + {\bf q^8t^3} +{\bf q^{10}t^3}-{\bf q^{10}t^4}\Big)
\cdot\frac{(1-{\bf q^{12}t^8})^{^\frac{n-\blue{3}}{4}}}{ {1-{\bf q^6t^4}}}
- \right.\nn \\
- \underbrace{{\bf q^{16}t^{10}}(1+{\bf q^6t})
\Big(1+{\bf q^2}   + {\bf q^6t}+{\bf q^6t^2} + {\bf q^8t }+{\bf q^{10}t^3}\Big)}_
{{\bf q^{18}t^{10}}(1+{\bf q^6t})\Big(\frac{1}{\bf q}[2](1+{\bf q^6t})
+\frac{1}{\bf q^2}{\bf q^6t^2}(1+{\bf q^4t})\Big)}
\left.
\cdot\frac{1-({\bf q^{16}t^{10}})^{^\frac{n- {3}}{4}}}{(1-{\bf q^4t^2})(1-{\bf q^{8}t^{4})}
}\right)
\nn
\ee
}

\noindent
Throwing away all the $n$-dependent powers we obtain the projective limit:
\be
_r {\cal K}^{[4,\infty]}_{_\Box}(N=3)
= \frac{ (1+{\bf q^8t^3})(1+{\bf q^{10}t^5})}
{(1-{\bf q^6t^4})(1-{\bf q^8t^6})}
\Big(1 +{\bf q^4t^2} +{\bf q^8t^4}\Big) =
\ee
{\footnotesize
\be
= 1 +\frac{\bf q^4t^2}{(1-{\bf q^6t^4})(1-{\bf q^8t^6})}\cdot\Big(
(1+{\bf q^2t})\cdot {\bf q^{12}t^5} + (1+{\bf q^4t})(1+[2]{\bf q^3t^2}+{\bf q^{14}t^7})
+ (1+{\bf q^6t})\cdot({\bf q^4t^4}(1-{\bf q^6t^4})+{\bf q^{16}t^9})
\Big)
\nn
\ee
}

\noindent
in full accordance with \cite{Gsl3}.
Since the expression in the first line is proportional to $(1+{\bf q^2t})$,
in {\it unreduced} case one could expect
{\footnotesize
\be
? \ \
 \frac{1}{\bf q^2}\frac{ (1+{\bf q^8t^3})(1+{\bf q^{6}t})}
{(1-{\bf q^6t^4})(1-{\bf q^8t^6})}
\Big(1 +{\bf q^4t^2} +{\bf q^8t^4}\Big)\cdot\frac{1+{\bf q^{10}t^5}}{1+{\bf q^2t}}
 \ \ ?
\nn
\ee
}
However,  this is not a positive series.
The true {\it unreduced} expression arises by application of the rules
\be
[3](1+{\bf q^2t})\longrightarrow \frac{1}{\bf q^2}(1+{\bf q^6t}), \ \ \ \ \
\begin{array}{ccccc}
&&\phantom.[3][2]{\bf q}(1+{\bf q^6t})\\
&\swarrow && \searrow \\
\phantom.[3](1+{\bf q^4t})\!\!\!\!\!\!\!\!\!\!\!\!\!\!\!\!\!\!\!\!
&&\longrightarrow &&
\!\!\!\!\!\!\!\!\!\!\!\!\!\!\!\!\!\!\!\! \frac{1}{\bf q}[2](1+{\bf q^6t})
\end{array}
\label{reduct3}
\ee
to decomposition in the second line:
\be
_{ur} {\cal K}^{[4,\infty]}_{_\Box}(N=3)
 = [3]+\frac{{\bf q^2t^2}(1+{\bf q^6t})\Big(
1+{\bf q}^2  + [2]^2{\bf q^4t^2}   +[3]{\bf q^6t^4}(1-{\bf q^6t^4})
 + {\bf q^{12}t^5}
+[2]{\bf q^{15}t^7}  +[3]{\bf q^{18}t^9}
\Big)}
{(1-{\bf q^6t^4})(1-{\bf q^8t^6})}
\nn
\ee
The same rules convert above expressions for the full {\it reduced} polynomials
${\cal K}^{[4,n]}$ into {\it unreduced} ones -- this is illustrated by expansions
of particular coefficients, not captured by projective limit.
In the product of two (or more) differentials reduction acts on the lowest one,
e.g., $(1+{\bf q^2t})(1+{\bf q^6t}) \longrightarrow \frac{1}{\bf q^2}(1+{\bf q^6t})^2$ --
otherwise we would get $[3](1+{\bf q^2t})(1+{\bf q^6t})$ which contains three
time more items.

Topological invariance implies that
\be
{\cal K}_r^{[4,3]}(N=3) = {\cal K}_r^{[3,4]}(N=3) = {\cal P}_r^{[3,4]}({\bf a}={\bf q}^3)
\ee
what is indeed the case.
For unreduced polynomials the analogue of the second relation is incorrect, because
unreduced KR polynomial always differs from unreduced superpolynomial,
even for $N=m=3$, see \cite{AnoMredP}.

\section{Remnant  of differential expansion}

Differential expansions (DE) of \cite{evo} and \cite{Arthdiff} play an increasingly
important role in modern theory of knot polynomials.
Essential for KR reductions will be a complementary {\it nested DE} \cite{AnoMredP}.
Somewhat amusingly, the ordinary DE partly survives the Khovanov reduction
at $N=2$ as well.
Namely, while
\be
{\cal P}_{_\Box}^{[m,n]} - 1 \ \vdots \ \{Aq\}\{A/t\} \sim
({\bf a^2q^2t^3})\cdot\left(1+\frac{\bf a^2t}{\bf q^2}\right)
\ \ \Longrightarrow  \ \
\nn \\
\Longrightarrow \ \ \
{\cal P}_{_\Box}^{[m,n]}({\bf a}={\bf q^2}) - 1 \ \vdots \
(1+{\bf q^2t})(1+{\bf q^6t^3})\sim
\ \ \Longrightarrow  \ \
{\cal P}_{_\Box}^{[m,n]}({\bf a}={\bf q^2}) - 1 \ \vdots \ (1+{\bf q^2t})^2
\label{PDE2}
\ee
for Khovanov polynomials we have:
\be
\boxed{
{\bf q}^{\gamma_{[m,n]}}\cdot {\cal K}_{_\Box}^{[m,n]}  - 1 \ \vdots \ (1+{\bf q^2t})
}
\label{DEK2}
\ee
but the power at the r.h.s. can not be raised from one to two.

We emphasize that the normalization of Khovanov polynomials is already fixed by the
condition (\ref{KRfroSUP}),
\be
{\cal K}_{_\Box}^{[m,n]}(N=2) - {\cal P}_{_\Box}^{[m,n]}({\bf a}={\bf q^2})
\ \vdots \ (1+{\bf t})
\ee
thus the mysterious power of ${\bf q}$ at the l.h.s. should have some objective meaning.
It actually depends on the number of strands $m$ and, as usual, on the residue
$r=n\, {\rm mod}\, m$:
{\footnotesize
\be
\!\!\!\!\!\!
\begin{array}{c||c||c|c||c|c||c|c|c|c||c|c||c||c}
m & 2 & 3 && 4 && 5 &&&& 6 && \ldots & m  \\
\hline
n & 1+2k& 1+3k& 2+3k & 1+4k& 3+4k & 1+5k & 2+5k& 3+5k& 4+5k & 1+6k& 5+6k && 1+mk   \\
\hline
\gamma_{[m,n]} & 0 & 2k & 2k & 4k& 2+4k& 8k&8k & 2+8k&4+8k& 12k & 8+12k &&
2k\cdot{\rm entier}\!\left(\frac{(m-1)^2}{4}\right)
\end{array}
\nn
\ee
}

\noindent
The generating function for the coefficients in front of $k$ is
$\sum_m {\rm entier}\!\left(\frac{(m-1)^2}{4}\right)\cdot x^{m} = \frac{x^2}{(1-x)^2(1-x^2)}$,
residue-dependence is easily restored from the topological identity
$\gamma_{[m,r]}=\gamma_{[r,m]}$.
Differential $d_1\sim\{A/t\}$,\footnote{
Strictly speaking, for ${\bf a=q^2}$ the factor $\{Aq\}$ is also {\it proportional}
to $d_1$ -- this is the origin of the square at the r.h.s. of (\ref{PDE2}) --
but most probably this factor is fully destroyed by the Khovanov reduction.
}
which is surviving in these formulas in the form
$(1+{\bf q^2t})$,
is responsible for the $U(1)$-reduction of knot polynomials,
i.e. the DE relation (\ref{DEK2})
implies that there
is some memory about the $U(1)$ reduction, surviving when that to $U(2)$  is performed,
but there is a nontrivial correction when ${\cal K}$ is not just ${\cal P}({\bf a=q^2})$,
i.e. when $\xi\neq 0$ in (\ref{xidef}).

For KR polynomials at $N=3$ -- the only beyond $N=2$ one available at the moment  --
the structure is more complicated
\be
{\bf q}^{?}\cdot{\cal K}_{_\Box}^{[m,n]}(N=3) - 1 \ \backslash\!\!\vdots \ (1+{\bf q^4t})
\ee
for any power of ${\bf q}$.
Still {\it some} structure clearly exists in this case as well --
and deserves closer attention.

\section{Conclusion}

In this paper we made six
claims:

\bigskip

{\bf (a)}
KR polynomials for torus knots
possess just the same $n$-evolution expansion as the superpolynomials:
\be
K_R^{[m,n]}(N,{\bf q},{\bf t}) = \sum_{Q\in R^{\otimes m}} k_Q(N,{\bf q},{\bf t})\cdot \lambda_Q^n
\label{Kevo}
\ee

\bigskip

{\bf (b)}
When quantum dimension $D_Q$ vanishes, so does the expansion coefficient $k_Q$:
\be
D_Q(N) = 0 \ \Longrightarrow  \ \ M_Q^*(N) \ \ \Longrightarrow \ \  k_Q(N) =0
\ee
and we get a {\it shortened} recursion, which needs less initial conditions.


\bigskip

{\bf (c)}
In variance with (\ref{evo}) for {\it reduced} torus superpolynomials,
the evolution expansion (\ref{Kevo}) for their KR counterparts
is not  consistent   with  
the mirror symmetry under $n\longleftrightarrow -n$.
Strictly speaking, the lack of covariance under $n\longleftrightarrow -n$
is not a problem, because the corresponding knot-diagrams are not Reidemeister equivalent.
Still the two knots, though topologically different, are related by a mirror map
and thus one could expect a relation like (\ref{invn}).
Moreover, if we had a control over these transformation laws, the knowledge of any
particular KR polynomial would provide {\it two} initial conditions.
For $m=2,3$ this {\it mirror anomaly} is concentrated in the framing factor
and thus is indeed comprehensible,
however it becomes far more complicated for $m\geq 4$. 
Technically it is related to the presence of negative items in the coefficients
in front of the fractions $\frac{1-\lambda^n}{1-\lambda}$.
Actually these negative items are absorbed into structures,
proportional to $(1-\lambda)$, and thus
do not spoil positivity of entire polynomial --
but this works only for $n>0$, while at $n<0$
such combinations fail to be sign-definite and can not pretend to be
super- and KR polynomials.
This is an unpleasant property of the evolution formulas,
already encountered in a similar situation in \cite{Sat}.

\bigskip

{\bf (d)} The knowledge of the evolution recursion in $n$ allows one to easily calculate
Khovanov polynomials with $N=2$ for arbitrary torus knots, beginning actually from the unknot(!):
one does it first for arbitrary $n$ at $m=2$ and then proceeds recursively in $m$,
using the topological identity between $[m,n]$ and $[n,m]$.
Important here is the claim (b): dramatic reduction of recursion degree,
which at  $N=2$ grows linearly with $m$,
what makes above identity sufficient for the $m$-recursion 

\bigskip

{\bf (e)}
Though differential expansion \cite{Arthdiff}, implying that
\be
{\cal P}_{_\Box}-1\sim\{Aq\}\{A/t\}
\ee
for fundamental HOMFLY and superpolynomials in topological framing,
is violated for KR polynomials, it is substituted by a weaker, still powerful
structure: a linear decomposition of properly  
normalized
$K_{_\Box}$
into differentials:
\be
{K}_{_\Box}(N)-1 = \oplus_{k=1}^N (1+{\bf q}^{2k}{\bf t})
\ee
As a simplest application,
reduction rule (\ref{redqn})  applies nicely
to Khovanov 
 polynomials and it allows to directly obtain
{\it unreduced} polynomials from {\it reduced} ones.
In result {\it unreduced} Khovanov 
 polynomial always satisfies
\be
_{ur}{K}_{_\Box}(2 
)-[2 
] \sim (1+{\bf q^4}{\bf t})
\ee
However, this kind of rule are {\it not} sufficient to further deduce
{\it reduced} Khovanov {\it per se} from superpolynomials and
more effort is needed for this purpose.
More relevant can be the nested structure, which
helps to fix the ambiguities in linear decomposition,--
but this story is beyond the scope of this letter.

\bigskip

{\bf (f)} A more traditional version of differential expansion is
found to survive for Khovanov  polynomials ($N=2$) in a somewhat surprising form of
(\ref{DEK2}), with additional powers of ${\bf q}$,
specific for torus knots.
It would be interesting to see what happens to these  relations
beyond the present scope -- of torus knot and fundamental representations.

%
%

\section*{Acknowledgements}

We are indebted to Eugene Gorsky for discussions, encouragement and sharing his
lists of torus KR polynomials for $N=2$ and $N=3$.
We also appreciate the  comments of Dror Bar-Natan and Lukas~Lewark.
We are grateful to Gleb~Aminov, Semeon~Arthamonov, and Alexander~Popolitov
for help with computing Khovanov polynomials
and to Alexander~Shumakovitch for sharing his program for reduced Khovanov homologies.

This work was supported by the Russian Science Foundation (Grant No.16-12-10344).

\end{document}